# Consensus Statement on Brillouin Light Scattering Microscopy of Biological Materials[§]

Pierre Bouvet[1], Carlo Bevilacqua[2], Yogeshwari Ambekar[3], Giuseppe Antonacci[4], Joshua Au[3], Silvia Caponi[5], Sophie Chagnon-Lessard[6,7], Juergen Czarske[8,9,10], Thomas Dehoux[11], Daniele Fioretto[12], Yujian Fu[1], Jochen Guck[13], Thorsten Hamann[14], Dag Heinemann[15], Torsten Jähnke[16], Hubert Jean-Ruel[7], Irina Kabakova[17], Kristie Koski[18], Nektarios Koukourakis[8,9], David Krause[8,9], Salvatore La Cavera III[19], Timm Landes[15], Jinhao Li[2], Jeremie Margueritat[11], Maurizio Mattarelli[12], Michael Monaghan[20], Darryl R. Overby[21], Fernando Perez-Cota[19], Emanuele Pontecorvo[22], Robert Prevedel[2,23], Giancarlo Ruocco[24], John Sandercock[25], Giuliano Scarcelli[3], Filippo Scarponi[25], Claudia Testi[24], Peter Török[26-28], Lucie Vovard[11], Wolfgang Weninger[1], Vladislav Yakovlev[29-31], Seok-Hyun Yun[32], Jitao Zhang[33], Francesca Palombo[34], Alberto Bilenca[35], Kareem Elsayad[1]

1. Center for Anatomy and Cell Biology, Medical University of Vienna, Austria
2. Cell Biology and Biophysics Unit, European Molecular Biology Laboratory, Germany
3. Fischell Department of Bioengineering, University of Maryland, USA
4. Specto Srl., Milan, Italy
5. CNR - Istituto Officina dei Materiali (IOM), Unità di Perugia, Italy
6. LightMachinery Inc., Canada
7. Departement of Electronics, Carleton University, Canada
8. Laboratory of Measurement and Sensor System Technique (MST), TU Dresden, Germany
9. Cluster of Excellence Physics of Life, TU Dresden, Germany
10. Competence Center for Biomedical Computational Laser Systems, TU Dresden, Germany
11. Institut Lumière Matière, UMR5306 Université Lyon 1-CNRS, Université de Lyon, France
12. Dipartimento di Fisica e Geologia, Università di Perugia, Italy
13. Max Planck Institute for the Science of Light, Erlangen, Germany
14. Department of Biology, Norwegian University of Science and Technology, Trondheim, Norway
15. Hannover Centre for Optical Technologies, Leibniz University Hannover, Germany
16. CellSense Technologies GmbH, Berlin, Germany
17. School of Mathematical and Physical Sciences, University of Technology Sydney, Australia
18. Department of Chemistry, University of California Davis, USA
19. Optics & Photonics Group, Faculty of Engineering, University of Nottingham, United Kingdom
20. Discipline of Mechanical, Manufacturing & Biomedical Engineering, Trinity College Dublin, Ireland
21. Department of Bioengineering, Imperial College London, United Kingdom
22. CREST Optics S.p.A., Rome, Italy
23. German Center for Lung Research (DZL), Heidelberg, Germany
24. Center for Life Nano- & Neuro-Science, Istituto Italiano di Tecnologia, Rome, Italy
25. The Table Stable Ltd., Mettmenstetten, Switzerland
26. School of Physical & Mathematical Sciences, Nanyang Technological University, Singapore
27. Lee Kong Chian School of Medicine, Singapore Centre of Environmental Life Sciences Engineering (SCELSE), Nanyang Technological University, Singapore
28. Institute for Digital Molecular Analytics & Sciences, Nanyang Technological University, Singapore
29. Department of Biomedical Engineering, Texas A&M University, USA
30. Department of Electrical and Computer Engineering, Texas A&M University, USA
31. Department of Physics and Astronomy, Texas A&M University, USA
32. Harvard Medical School and Massachusetts General Hospital, USA
33. Department of Biomedical Engineering, Wayne State University, USA
34. Department of Physics and Astronomy, University of Exeter, Exeter, United Kingdom
35. Biomedical Engineering Department, Ben-Gurion University of the Negev, Israel

\* Correspondences can be addressed to: kareem.elsayad@meduniwien.ac.at

---

**Brillouin Light Scattering (BLS) spectroscopy is a non-invasive, non-contact, label-free optical technique that can provide information on the mechanical properties of a material on the sub-micron scale. Over the last decade it has seen increased applications in the life sciences, driven by the observed significance of mechanical properties in biological processes, the realization of more sensitive BLS spectrometers and its extension to an imaging modality. As with other spectroscopic techniques, BLS measurements not only detect signals characteristic of the investigated sample, but also of the experimental apparatus, and can be significantly affected by measurement conditions. The aim of this consensus statement is to improve the comparability of BLS studies by providing reporting recommendations for the measured parameters and detailing common artifacts. Given that most BLS studies of biological matter are still at proof-of-concept stages and use different--often self-built--spectrometers, a consensus statement is particularly timely to assure unified advancement.**

In the context of studying bio-relevant matter, BLS spectroscopy involves measuring the frequency and lifetime of MHz-GHz frequency acoustic phonons (propagating density fluctuations) in a material. This is possible by making use of the constructive and destructive interference of light scattered from propagating thermal or stimulated phonons of a given wavevector determined by the Bragg condition[1]. A BLS spectrum of a homogeneous and isotropic material will consist of at least two (Stokes and anti-Stokes) peaks that reveal the amplitude, frequency and lifetime of acoustic phonons at the probed scattering wavevector(s).

Examples of the BLS spectra of distilled water, measured using different spectrometer designs, are shown in Figures 1A-E. Also shown in each case is the spectrum of (>99% pure) cyclohexane, which has a smaller BLS frequency shift ($\nu_B$) and broader linewidth ($\Gamma_B$) than water. A description of the different spectrometer designs can be found in the Supplementary Text. In each case, the frequency shift of the BLS peak corresponds to the probed phonon frequency. Using the wave equation this can be used to calculate the phonon hypersonic speed, which is related to the respective elastic modulus (see Supplementary Text). The width of the BLS peaks is related to the imaginary (dissipative) part of the modulus which can be expressed in terms of the longitudinal viscosity, although it is also dependent on extrinsic experimental conditions and is in general affected by spatial heterogeneities in the sample in a non-trivial manner.

While BLS spectroscopy has been performed in research laboratories for more than half a century[2], the recent surge of interest for biomedical applications can largely be attributed to its extension to an imaging modality[3,4] suitable for studying biological samples[4]. Driven also by our growing awareness of the significance of mechanical properties for biological processes[5,6], the last decades has seen the development of a number of different approaches[7,8] to obtain BLS spectra of biological matter (both in the frequency and in the time domain, and both from inherent thermal and stimulated acoustic phonons), that have been applied for studying diverse biological and biorelevant systems[8-10]. These, in principle, are expected to yield comparable results for the viscoelastic moduli. Each approach however has distinct experimental challenges, limitations and characteristic sources of uncertainty. There are ongoing discussions on the biological significance of BLS-derived parameters in the context of soft tissue and cells[11-15], here though we focus on the optics and photonics aspects necessary to extract reliable and repeatable parameters in biological matter. We also only consider BLS performed at visible to NIR wavelengths, as X-ray[16] and neutron[17] Brillouin scattering are associated with distinct challenges.

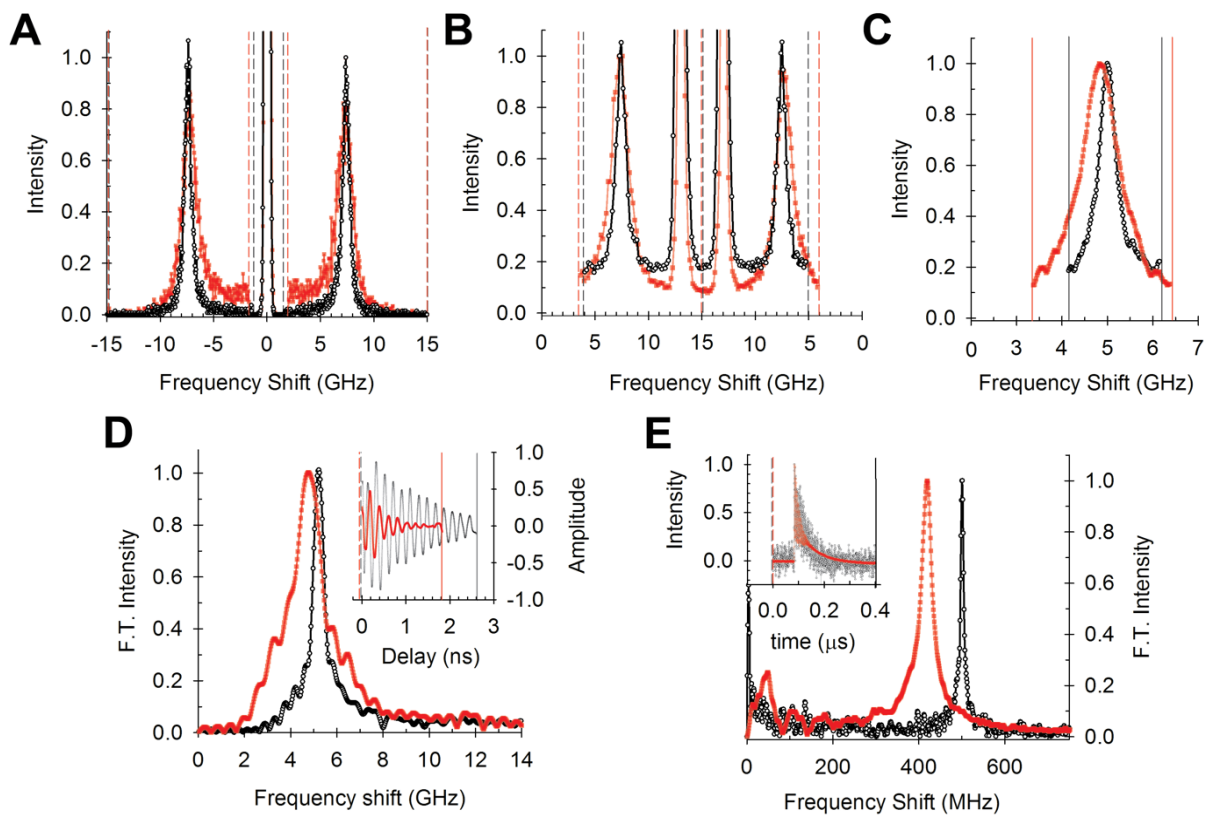

**Figure 1: Brillouin Light Scattering Spectra of Liquids.** Example BLS spectra of distilled water (black circles) and cyclohexane (red squares) measured using different spectrometer designs (normalized to unity). **A** Tandem Fabry Perot (TFP) spectrometer, **B** Virtual Imaged Phase Array (VIPA) spectrometer with Electro-Optic Modulator (EOM) reference peaks at 13 GHz, **C** Stimulated Brillouin Scattering (SBS) spectroscopy, **D** Time Resolved Brillouin Scattering (TRBS) microscopy and **E** Impulsive SBS (I-SBS) spectroscopy. Insets of D and E show the respective measured time resolved data. The more pronounced difference between water and cyclohexane in E compared to A-D is due to the BLS frequency shift in I-SBS being independent of the refractive index.

# Main Text

## Parameters that BLS measures

Two *key parameters* measured in BLS spectroscopy are the BLS frequency peak shift ($\nu_B$) and the BLS linewidth ($\Gamma_B$), both conventionally presented in units of MHz or GHz. To obtain these parameters, one typically fits each Brillouin peak in the frequency domain with a suitable function (Supplementary Text).

Experimental data for a BLS measurement typically consists of the intensity at different frequencies that spans one or both Stokes and anti-Stokes BLS peaks, or time-series data longer than $1/\Gamma_B$ with a sampling frequency $>2\nu_B$. In most cases, BLS measurements do not measure the elastic scattering peak, which is often suppressed, masked or otherwise avoided, due to its large intensity. Below we list some physical quantities that can be extracted from the key parameters (equations for calculating these and others are provided in the Supplementary Text).

***Hypersonic acoustic speed (V):*** The phase velocity of the probed acoustic modes in the direction of the scattering wavevector. It is obtained from $\nu_B$, and depending upon the scattering geometry, the refractive index (*n*) of the sample in the direction of the scattering wavevector.

***Longitudinal Storage Modulus ($M'$):*** A measure of the elastic properties subject to longitudinal boundary conditions. For isotropic materials it is the combination of shear ($G'$) and bulk ($K'$) moduli: $M' = K' + (4/3)G'$ (Supplementary Text). In addition to all parameters required for calculating $V$, it also requires knowledge of the mass density ($\rho$).

***Longitudinal Loss Modulus ($M''$) and Longitudinal Viscosity ($\eta_L$):*** A measure of the viscous (dissipative) properties subject to longitudinal boundary conditions. In addition to all parameters required for calculating $M'$, it also requires $\Gamma_B$.

***Longitudinal Loss Tangent (*$\tan \delta$*):*** Defined as $M''/M'$, which is equal to $\Gamma_B/\nu_B$, a measure of attenuation independent of $\rho$, and usually to a good approximation *n*.

***Shear Modulus ($G$):*** Can be obtained from the BLS frequency shift of the Transverse Acoustic (TA) modes in a symmetry direction. TA phonon modes are manifested as two (Stokes and anti-Stokes) distinct peaks to those of Longitudinal Acoustic (LA) phonons, usually with a smaller BLS frequency shift. They typically require measurements at different scattering geometries and polarizations[18-21]. In soft-matter and liquids the frequency shift of the TA phonons may be practically too small to measure. Calculation of $G$ also requires $\rho$ and *n* in the direction of the scattering wave-vector. $G$ is a complex quantity, with the real and imaginary part obtainable from the BLS frequency shift and linewidth in an analogous manner to $M'$ and $M''$.

***Tensile or Young's moduli (E), Bulk Moduli (K), and Poisson's ratios ($\sigma$).*** These can be derived from measurements of *G* and *M* and require *a priori* knowledge or assumptions on the symmetry of the sample[18-20]. This is usually practical only in hard matter where the frequency of the TA phonon peak is not too small.

***Landau-Placzek Ratio***: Is defined as the ratio of the integrated Rayleigh scattering peak intensity (component of elastic scattering from the thermal diffusivity mode[22]) when measured, to the integrated BLS peak intensity (defined as the total area underneath the respective peak)[1]. For simple homogeneous samples it is equal to the difference between the specific heat at constant pressure and that at constant volume, relative to that at constant volume[1]. This can give insight into the thermodynamic properties of a sample, albeit the validity breaks down in complex materials and an accurate measure of the Rayleigh signal can be challenging in the presence of other elastic scattering processes (e.g. Tyndall scattering[23]). As such, its consideration in complex biological matter has thus far been avoided.

***Refractive Index ($n$)***: Can be calculated for isotropic materials or materials with known symmetry from angle-resolved measurements or from several different scattering geometries[19,24], under the assumption of a homogeneous sample, and at frequencies not in the vicinity of a structural relaxation process or phase change[25].

***Mass Density ($\rho$)***: Can be calculated in stimulated BLS measurements via the BLS gain, with which it scales linearly[26].

As can be seen, calculation of viscoelastic parameters typically require knowledge of extrinsic parameters ($\rho$ and *n*), which may not always be accessible. In homogeneous materials, $n$ can be measured using an Abbe refractometer, whereas in thin heterogeneous samples it can be spatially mapped using phase-resolved methods (quantitative phase imaging[27] or digital holographic tomography[28]). $\rho$ can be estimated from *n* via the Lorentz-Lorenz relation and knowledge of the proportionality coefficient(s)[29] (Supplementary Text). This however may not be trivial in heterogeneous samples, where these vary spatially. $\rho$ may alternatively be obtained using the linear gradient method[30] and for homogeneous samples using a densiometer[31] or measurement of sample mass and volume. If $\rho$ and $n$ are not accessible, then an approximation may be made from literature values. For samples with a spatially uniform chemical composition and material phase, a near constant ratio of $\rho/n^2$ may be assumed[32], however this can break down when samples have gradients in composition (e.g. interfaces between hydrophilic-hydrophobic regions).

## Resolution

While the spectral resolution of the spectrometer is a limiting factor, spectral deconvolution[33] and spectral fitting[34] can be employed for highly precise measurements of BLS frequency shifts and linewidths (exceeding the accuracy with which $v_B$ can be extracted by two orders of magnitude compared to the spectral resolution of the spectrometer[35]). For an unbiased comparison of the spectral precision of different BLS spectrometer setups the Allan variance method, introduced for IR spectroscopy, has been suggested[36,37]. This can be applied to BLS experiments by measuring standard deviations of the fitting parameters over the sampling time[38,39]. In order to obtain unbiased comparisons though, one should bear in mind that spectrometers may allow additional optimization by modulating the mirror spacings, as well as the acquisition/scan times, amplitudes, and number of acquisition points.

The spatial resolution is fundamentally limited by the effective Point Spread Function (PSF) of the microscope itself (with the exception of the axial resolution in Time Resolved Brillouin Scattering (TRBS)–see Supplementary Text). Given BLS is a coherent phenomenon, the effective PSF may be larger than the PSF defined using incoherent probes (e.g. fluorescent labels) or scattering processes. This can be understood in that the probed photon-phonon interactions are dependent on the coherence lengths of the phonons (i.e. the characteristic size over which the density fluctuations are correlated)[40,41]. This is sample dependent and given by $\Lambda \nu_B / \Gamma_B$, where $\Lambda$ is the phonon wavelength[42]. Increasing the mechanical contrast between different regions typically reduces the phonon coherence and, therefore, improves the spatial resolution up to the limit of the optical resolution obtainable from incoherent scattering processes. In practice, the true spatial resolution is best determined experimentally on mock-up systems, consisting of different materials with sharp interfaces and spatially scanning between these.

## Reporting Consensus

Below we describe parameters we consider important to report in all BLS measurements, regardless of the spectrometer used. These are provided also in the form of a *Minimal Reporting Table* (available at: DOI:10.6084/m9.figshare.27794910)

(1) **Spectrometer.** In addition to information on the type of spectrometer, this should also include its Free Spectral Range (FSR), sampling step size and range (in frequency for frequency domain or time for time domain).

(2) **Spectral resolution.** In frequency-domain spectrometers this is often most easily obtained from measuring the elastic scattering peak width from a highly scattering sample (e.g. mirror flat), with the assumption that the probing laser(s) have much narrower spectral line widths than the frequency shifts measured. In time-domain measurements this can be estimated from the repetition rate of the pulsed laser, which acts as the frequency comb spacing in the spectral domain. The spectral resolution is generally reported in units of MHz. While the spectral resolution is not a direct indication of the precision with which the key parameters can be extracted if the shape of the peak is known, it is important for distinguishing multiple peaks (which may occur near interfaces[43]) or deducing undefined/uncertain functional peak shapes (which may occur near structural transitions).

(3) **Spectral precision.** This is distinct to the spectral resolution in that it is a measure of how precisely $v_B$ and $\Gamma_B$ can be determined from repeated measurements, and ultimately a measure of the stability of the microscope/spectrometer. A reasonable practical measure of this can be obtained from the spread of repeated measurements or a spatial scan of a static homogeneous sample. In certain cases the difference in the absolute frequency shift and linewidth of the Stokes and anti-Stokes peaks may also contribute and should be considered. These parameters are generally reported as coefficients of variation on the mean (ratio of the standard deviation on the mean by the mean) as percentages (%). For a heterogeneous or time-varying sample the precision can have inherent spatio-temporal variability that is not accounted for with a single spectral precision, which should be considered. The acceptable spectral precision will depend on the size of the changes in the key parameters one wishes to discern. In practice a ~0.5% precision for $v_B$ and $\Gamma_B$ often suffices for discerning significant changes in the derived elastic and viscous parameters of biological samples.

(4) **Signal to Noise Ratio (SNR) and Signal to Background Ratio (SBR).** These are measures of how much the BLS signal compares to all other non-BLS detected signals. Various sources of background signals can exist including residual Amplified Spontaneous Emission (ASE), side modes of laser sources that are not perfectly clean, fluorescence signals, and artefacts from elastic scattering (e.g. from turbid samples like tissue or bones). The most desirable operating conditions are that all non-BLS signals are removed and that enough BLS photons are collected to operate in the shot noise limit, considering also the detector used. Reporting the SNR and SBR is therefore important to demonstrate if the measurement is shot noise (highly preferred) or background limited. Here the SNR may be quantified in an analogous manner to what has been used in Raman spectroscopy[44], as the ratio of the averaged BLS peak intensity to its standard deviation, while the SBR is the ratio of the averaged Brillouin peak intensity to the standard deviation of the noise floor in the absence of the Brillouin signal.

(5) **Numerical aperture (NA) of excitation and detection light / Scattering angle(s).** Since $v_B$ and $\Gamma_B$ are dependent on the scattering angle, using high NA's (>0.4) will result in broadening of the linewidth due to the increased spread of scattering wavevectors measured[45,46]. At scattering angles away from backscattering the effect of finite-NA becomes particularly pronounced and an experiment specific trade-off needs to be made on acceptable spread in scattering wave-vector, spatial resolution, laser exposure and acquisition time. While conventional wisdom states that increasing the excitation and/or detection NA results in a higher lateral spatial resolution, in BLS microscopy this may not always be the case on account of the finite phonon length scales[42]. As such, explicitly reporting both the excitation and detection NA or angles is essential for meaningful interpretation of data. In addition, for confocal implementations, reporting of the effective size of the detection pinhole (physical pinhole or fibre core diameter), *e.g.* in Airy Units, is important for understanding the confocality as well as the probed wavevectors.

(6) **Average Power/Peak Power at Sample/Pulse repetition rate.** Measurements conducted at low average laser power have the advantage of reduced photodamage/phototoxicity for live cell and tissue studies. They also reduce the potential of significant sample heating which affects the BLS spectrum. While for biological samples it is almost always desirable to use as low of a laser power as possible while still having an acceptable SNR, for dynamic/moving samples there may be a trade-off to capture the desired dynamics. In regard to not perturbing sensitive biological processes the total laser exposure (energy deposited on sample) is often more relevant, and as such a balance between exposure time and laser power may need to be found. The acceptable power-intensity is ultimately sample dependent and guidelines set out for other microscopy techniques should be followed[47].

(7) **Laser wavelength(s).** $\nu_B$ and $\Gamma_B$ depend on the wavelength of the laser source. When selecting wavelengths, both the strength of the BLS signal, the potential for photodamage as well as how deep inside a sample one wishes to measure, should be considered. While the BLS scattering cross-section scales as $\lambda^{-4}$ and the spatial resolution of a measurement increases with decreasing $\lambda$, the photodamage to biological materials is generally more significant at shorter wavelengths.

(8) **Exposure/Acquisition Time.** While the exposure time per spectral acquisition should be chosen to obtain sufficient SNR, dynamic/moving samples may set a practical upper limit, in which case a compromise of increasing laser intensity and decreasing spectral resolution may be needed.

(9) **Sample Temperature.** Given the strong temperature dependence of BLS, we recommend sample temperature should be reported to within +/-0.5°C accuracy. This will typically cause errors in the value of $\nu_B$ in hydrated samples of less than ≈0.1%.

## Methods

A description of the different approaches for performing BLS spectroscopy, together with measurement and reporting recommendations particular to each, are presented in the Supplementary Text. Here we discuss aspects relevant to different spectrometer designs, as well as the comparison of measurements on different setups.

## Calibration spectra

BLS spectra need accurate and robust calibration for data to be comparable across labs, experimental conditions, and over time. For instruments needing calibration using a reference inelastic peak, such as Virtually Imaged Phased Array (VIPA) architectures or single etalon systems, optimal performance is achieved by eliminating the elastic scattering, and calibration is typically performed using materials with known $\nu_B$ to estimate the unknown dispersion parameters. The accuracy of these calibration procedures rely on the accuracy of the reference $\nu_B$, which may depend on other experimental parameters such as temperature (e.g. $\nu_B$ of water increases by ~1%/°C; whereas $\nu_B$ of polystyrene decreases with increasing temperature[48]).

Calibration based on the absolute, and highly conserved, absorption lines of atomic gas vapours such as rubidium may be employed. It has been shown that locking the laser frequency to such narrow absorption lines allows for calibrated measurements that are accurate to within a few MHz over extended periods[49]. Alternatively, a robust calibration strategy can involve spectrally shifting a part of the elastic scattered signal by a few GHz using an electro-optic modulator (EOM)[50-52] or potentially also an acousto-optic modulator (AOM).

**Fitting of BLS spectra**

In cases where the presence of a mechanical relaxation process can be ruled out, the BLS peak is well represented by a Damped Harmonic Oscillator (DHO). When spectra are measured using Fabry-Perot interferometers or monochromators, a direct DHO fit is feasible. However, for imaging spectrometers where a spectral projection is generated by introducing a small divergence to the light incident on an interferometric element, and the signal is dispersed into one or more orders, the spectral projection will be weighted by a Gaussian envelope[53]. This results in a slight functional correction, and fits are typically performed using Lorentzian functions. Since the parameter $\nu_B$ in a DHO fit does not exactly coincide with the peak maximum a, usually small, correction may need to be applied for broad BLS-peaks when using Lorentzian fitting[54].

In the ideal scenario, the detected signal is shot noise limited, such that the variance $\sigma$ scales as $\sigma_i^2 = N_i$, where $N_i$ is the number of photons detected in a given frequency interval. The large variation in the relative photon count uncertainty across the BLS peak thus requires a weighted ($\propto \sigma_i^{-2}$) regression routine for fitting. In multi-pixel detectors there may also be an additional source of noise, e.g. due to electron amplification, that increases the variance, which should be considered. The fitting function may thus also include a baseline term to account for this and other physical phenomena (e.g. fluorescence and Raman scattering *folded* into the spectral region of interest due to the finite FSR of the spectrometer). This contribution and its uncertainty should be considered in the computation of the SNR. Rather than subtracting such contributions from the spectra, which can be statistically misleading, they should explicitly be included in the fitting function.

The measured BLS spectrum is always subject to instrumental broadening and should thus ideally be deconvolved with the (spectral) Instrument Response Function (IRF). Alternatively, a modified fitting function can be constructed by convolving the model function (DHO or Lorentzian) with the IRF. In frequency domain spectroscopy the IRF can be obtained by measuring the spectra of the elastically scattered probing laser through the optical setup by replacing the sample with a mirror flat in the backscattering geometry.

## Uncertainties

When reporting the uncertainties of a BLS measurement, it is recommended to follow the guidelines established by the international organisations of metrology[55]. Whenever possible, repeated measurements of the sample under the same conditions should be conducted to evaluate the uncertainty caused by random error sources. This type of uncertainty (Type A evaluation) is expressed as the standard deviation of the mean. For uncertainties that cannot be estimated from repeated measurements (Type B evaluation), such as systematic errors, a comparison with the reported values of standard materials should be performed when possible. The combined uncertainty can then be calculated by considering the dependencies between error sources. To mitigate systematic errors, regular calibration with readily available homogeneous materials with constant $v_B$ values, and where the BLS peak is well described by a model fitting function, is recommended. To report the uncertainty of any calculated viscoelastic parameters, the uncertainties of each of the measured or assumed parameters should be evaluated separately and then combined according to the law of propagation of uncertainty.

## Parameters for high-quality BLS measurements

One can distinguish between parameters that concern the illumination (probing) light and the detection of the BLS signal. On the illumination side, the most important parameters are: (1) *laser linewidth*. This determines the upper limit of the achievable spectral resolution, and should typically be <100 MHz. (2) *Long term laser wavelength stability*. This should ideally be <1GHz drift per hour. Locking the laser wavelength to a gas absorption cell line or re-calibration with control samples may be employed. (3) *Spectral noise of the laser* (amplified spontaneous emission--ASE and "side modes") can lead to considerable spectral contributions either side of the main laser line. ASE influences the spectral noise floor from its elastic scattering by the sample or within the optical setup. It can, on account of its spectral breadth, be challenging to filter out, potentially even requiring a combination of etalon filters. Typically, the total power outside of the main laser line should be ~60-90dB lower than that of the main laser line for high-quality BLS measurements.

On the detection side, it is most important to sufficiently suppress elastic scattered light. The required magnitude thereof depends on a sample's scattering properties, although a suppression of ~60-90dB (less in homogenous, low-scattering samples such as transparent liquids, more in highly scattering samples such as calcified bones) is desirable. At the same time, the detection of the BLS signal needs a sufficiently high SNR for high-precision fitting. If a (typical) precision of ~10MHz is desired, this requires an SNR of between ~30-50, depending slightly on other spectrometer properties and acquisition parameters (spectral sampling, etc.)[56]. Given that state-of-the-art spectrometers are often shot noise limited, this would correspond to a total signal that is on the order of $10^3 - 10^4$ detected photons across the measured BLS spectrum.

While a high SNR, spectral resolution and precision will result in more reliable determination of the key parameters, how accurately they can be extracted ultimately depends on how accurately the fitting model describes the measured spectra. For this including finite-NA peak broadening, multiple scattering in highly scattering samples, background noise, and deconvolving with the IRF will all play a part.

## Artifacts

Different spectrometer designs will have unique sources of artifacts, which are described in the Supplementary Text. Below we list artefacts that were identified as being common to several spectrometer designs, and which can affect determination of the BLS key parameters.

**Material (acoustic) heterogeneities:** When elastic heterogeneities exist in the scattering volume, the BLS spectrum may no longer exhibit a single set of peaks (per acoustic mode). While not necessarily an artefact, since it can provide insight into the composition of a sample, it nevertheless may require explicit consideration. The effect will depend on the size and the acoustic mismatch of the heterogeneities[40,41,57]. When the size of the acoustic heterogeneities is larger than the phonon coherence length, one will have distinct peaks associated with the different materials that can be analyzed independently[58] (with an intensity proportional to the product of the filling factor and scattering efficiency[57]). If the size of the heterogeneities is less than the phonon coherence length and there is a significant acoustic mismatch associated with the heterogeneities, the acoustic modes will be confined, and the BLS spectra will exhibit peaks whose position is a function of their morphological and acoustic properties. Here a specific analysis procedure is required[41,59,60]. Finally, if the phonon coherence length is larger than the heterogeneities, one may assume an effective material response. Here the heterogeneities cause a *static* attenuation process, which adds to the inherent dynamic ones, and is associated with an increase in $\Gamma_B$[40].

**Scattering wavevector indeterminacy:** An indeterminacy in the probed scattering wavevector ($q$) will be reflected in both $\nu_B$ and $\Gamma_B$, and any subsequently derived parameters. Given that the probing and collection optics have a finite NA, this effect is unavoidable. In the case of low NA and in the back scattering configuration this effect is largely reduced[46]. However for high NA excitation or detection, as well as for small scattering angles[61], the effect on both $\nu_B$ and $\Gamma_B$ can become very significant[42,45]. Given that the theoretical evaluation of the $q$-indeterminacy may not be straightforward, it can be desirable to estimate this from measurements on materials with negligible intrinsic broadening, which can serve to also obtain the IRF of the optical setup.

In turbid media where one has significant multiple scattering, the effects of $q$-indeterminacy becomes more complicated, as the probed phonons are no longer defined by the external scattering geometry. Here the observed spectrum is a complex superposition of BLS spectra with different scattering geometries. Methods based on e.g. polarization gating[62] may be used to separate different contributions and extract the true BLS parameters.

**Sample heating effects:** Temperature has a significant effect on the BLS key parameters as well as cellular processes. In hydrated biological samples an increase in temperature will typically result in an increase of $\nu_B$ and decrease of $\Gamma_B$. As such it is important to consider uncertainties in the key parameters due to uncertainties in the temperature. To check whether local sample heating is significant, it can suffice to perform equivalent measurements with different laser intensities or sample exposure times.

Given that some BLS approaches require intense and/or extended sample laser exposures, sample heating may become relevant. In general, this is non-trivial to calculate or measure in-situ and will depend upon numerous parameters. Semi-quantitative estimates can be made using *e.g.* the bio-heat equation[63].

In time-domain BLS the temperature rise at the transducer caused by the absorption of laser pulses may also contribute to sample heating. This is most significant when using large average laser powers and thermally insulating substrates. Using a sapphire substrate with 1.5 mW average laser power, it was shown that one would have a ~1°C temperature rise in the sample measurement volume[64].

**Laser frequency drifting or line-broadening:** Ideal lasers for Brillouin scattering are narrow linewidth (< 10MHz) at a single, stable frequency. Some commercial continuous wave (CW) single-longitudinal mode lasers, such as those commonly used in spontaneous and stimulated frequency-domain BLS, experience frequency drifts, or line-broadening, that are predominantly caused by temperature fluctuations. Such frequency drifts can be >100 MHz for environmental temperature variations of 0.1°C, although the amount will vary depending on the laser type and manufacturer[65]. To increase wavelength stability during BLS spectra acquisitions, laser sources may be actively locked to a reference (*e.g.* Rb/I$_2$ gas cell, or a stable external cavity). Depending on the time scales associated with frequency drifts, one may perform periodic measurements of a reference material or signal with known $\nu_B$ and implement frequency drift corrections during analysis.

**Elastic scattering filters:** Given the tight frequency-matching requirements, stability represents the main challenge for most elastic scattering filters. The suppression of the elastically scattered light can be compromised by laser frequency drifts, mechanical drifts, environmental thermal fluctuations, as well as intrinsic Amplified Spontaneous Emission (ASE) from the laser. As such, the use of laser lock-in or closed loop configurations that involve a continuous readout of the elastic scattering signal, are generally required for both interference and molecular absorption-based filters. For the latter, the multitude of absorption lines (especially in I$_2$ absorption cells) may attenuate or distort the BLS peaks, which may need to be considered during analysis.

**Spectral overlaps:** BLS spectrometers operating on interferometric principles will have a finite FSR. In measuring a sample that has a $\nu_B$ larger than the FSR, the BLS peak(s) will appear in higher orders or overlapped orders, if the spectrometer is not explicitly designed to cancel out overlapping and higher order peaks[66]. For example, if a sample has $\nu_B = 45$ GHz yet the spectrometer FSR is 30 GHz, the BLS anti-Stokes peak will deceptively show up at 15 GHz. This ambiguity, and the measurement of $\nu_B$ values larger than the FSR of a spectrometer, can be overcome by measuring the BLS peaks in higher order ranges[67]. In dual etalon TFP setups such spectral overlaps are eliminated by selecting different etalon FSRs[68]. The issue may also be mitigated in dual-VIPA setups by using VIPA's with different FSRs effectively increasing the FSR of the spectrometer[69]. This ambiguity in $\nu_B$ will not be an issue in stimulated and time-domain BLS.

**Internal reflections in sample:** For samples that are thin, highly reflective, or support optical waveguiding modes, multiple reflections from interior surfaces can give rise to additional BLS peaks in non-backscattering geometries[70,71]. This is due to light reflected in the interior of a sample leading to two (or more) inner scattering angles. The result is the observation of two (or more) sets of BLS peaks, one associated with the experimentally chosen scattering geometry and the other corresponding to backscattering from internal reflections. The (longitudinal) modes with the largest $\nu_B$ will generally be those associated with backscattering, and may show angular dependence with sample rotation[70]. Since transverse modes are forbidden in backscattering for high symmetry conditions, only longitudinal modes will appear from such internal scattering and can be identified by employing crossed polarizers in the probing and detection beam paths.

To illustrate some potential artifacts that can occur in BLS microscopy, in Fig 2A-D we show spatial maps of the BLS frequency shift and linewidth for a 10 μm diameter polystyrene bead measured in four different laboratories. Experimental details can be found in the respective *Minimum Reporting Tables* (available at DOI:10.6084/m9.figshare.27794916). In Fig 2A and 2B the bead has been embedded and sealed in pure (>99%) glycerol which has a larger BLS frequency shift than the bead. In Fig 2C and 2D the embedding medium is glycerol exposed to air for ~24 hours, and an aqueous solution respectively, whose BLS frequency shift is less than that of the bead. These two scenarios (Fig 2A and 2B versus Fig 2C and 2C) create distinct acoustic and optical interfaces between the bead and its surroundings, which can lead to distinct imaging artifacts.

The significant increase in the linewidth at the edge of the bead in Figures 2A and 2B can be partly attributed to the probing volume containing both the bead and the surrounding media (and thus contributions from phonons in both materials), resulting in what is effectively a broader peak. In addition, the acoustic impedance mismatch presented by the curvature of the bead surface may result in an increased spread of scattering angles away from true backscattering. These are hard to avoid given the spherical geometry of the bead, and can also explain the intermediate values of the BLS frequency shift observed at the bead surface in Figure 2B. Abrupt acoustic and optical interfaces generally present a challenge in BLS microscopy, often best addressed by fitting two Stokes and two anti-Stokes BLS peaks in their vicinity.

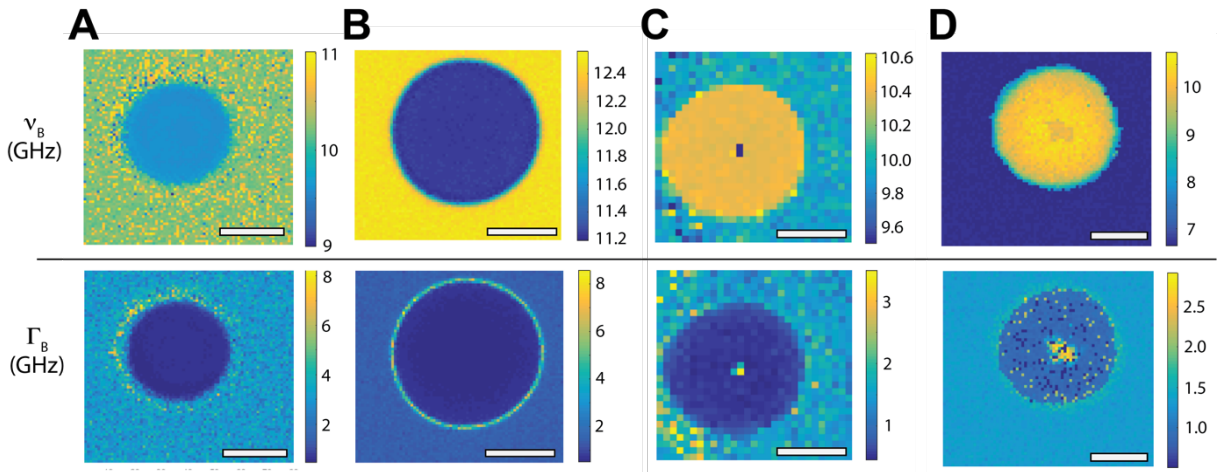

**Figure 2: Spatial maps of BLS frequency shift ($\nu_B$) and linewidth ($\Gamma_B$) for a 10µm polystyrene bead. A:** TRBS microscopy measurement with 758nm probing wavelength. **B-D** VIPA-spectrometer measurements with 660nm probing wavelength. A & B were embedded in glycerol (>99%), whereas C and D were measured in glycerol with a significant water fraction and Optiprep[TM] density gradient medium respectively, that both have a lower acoustic index than the bead. Scale bars = 5µm. *Minimum Reporting Tables* for measurements can be found at DOI:10.6084/m9.figshare.27794916.

The region of decreased $\nu_B$ and increased $\Gamma_B$ seen at the center of the beads in Figures 2C and 2D, can be understood as a geometrical artifact of the bead acting like a lens, as would occur when the bead is immersed in a lower refractive index media. Measurements at this central position would as a result be over a broader range of also smaller scattering wavevectors (effectively decreasing $\nu_B$ and increasing $\Gamma_B$ as seen). The bead represents an extreme example due to the atypically large refractive index and acoustic impedance mismatch with its surroundings (rarely encountered in soft biological matter), but can serve as a useful illustration of potential geometric and material artifacts.

## Comparison of absolute and relative BLS measured values

Figures 3A and 3B show the acoustic speed ($V = 2\pi\nu_B q^{-1}$) and kinematic longitudinal viscosity ($\mu_L = 2\pi\Gamma_B q^{-2}$) of distilled water obtained from measurements on different spectrometers in 15 independent laboratories around the world. Despite measuring at different wavevectors and scattering geometries, all results should yield the same values of $V$ and $\mu_L$. Values of the $\nu_B$ derived $V$ are in good agreement between laboratories and spectrometer designs (typically within <0.5%). The $\Gamma_B$ derived $\mu_L$, while all having similar trends with respect to temperature, show a much larger variability. This is likely due to numerous compounding effects that need to be addressed on a case-per-case basis, including lack of proper spectral deconvolution, finite time windows used to generate spectra (for I-SBS and TRBS), and inadequate corrections for finite probing/detection NA. Among a given spectrometer design the latter presumably gives the largest contribution to the observed variability, since the combined linear and quadratic wavevector dependence of $\nu_B$ and $\Gamma_B$ means that even a small change in NA can result in a significant change in the measured $\Gamma_B$.

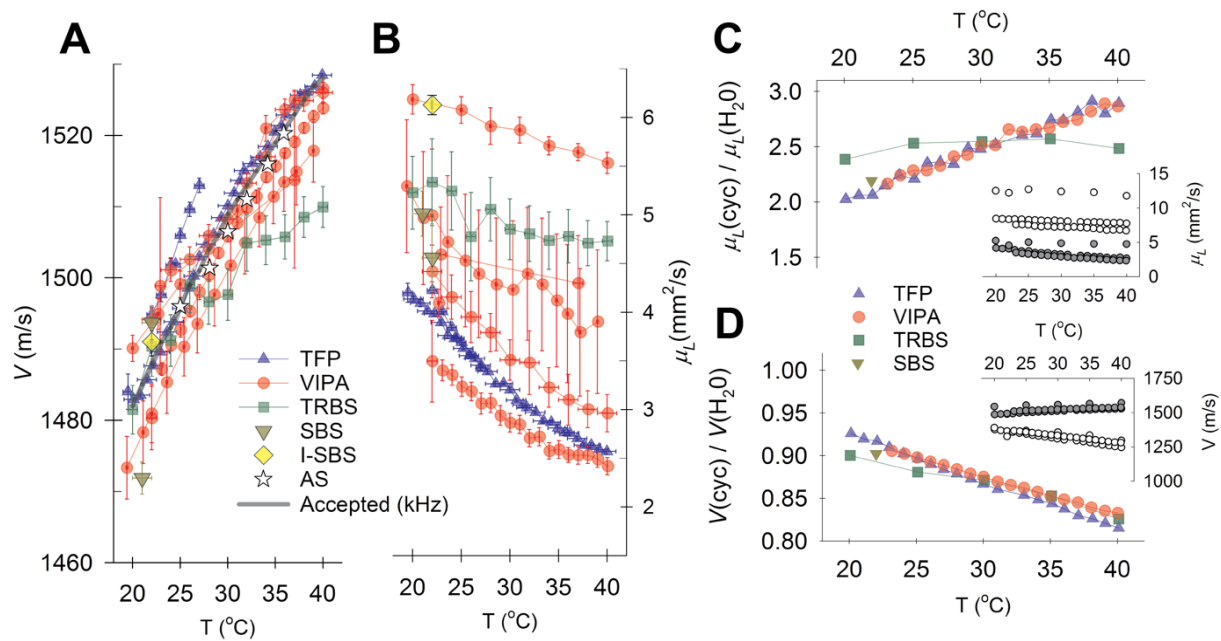

**Figure 3: Comparison of Brillouin Light Scattering derived parameters from different spectrometers. A & B** Hypersonic speed ($V$) and kinematic longitudinal viscosity ($\mu_L$) as a function of temperature measured using different spectrometer designs in participating laboratories. Also shown are accepted values for the acoustic speed in water, and values reported from Acoustic Spectroscopy (AS) measurements[72]. **C & D** BLS obtained $\mu_L$ and $V$ of cyclohexane relative to that of pure water measured using different BLS techniques after normalization to Vienna-Hanover water standard--see *Supplementary Text*. Insets show absolute values for water (filled circles) and cyclohexane (open circles).

Despite the large variability in the measured $\mu_L$, it is possible to obtain comparable absolute values of $\mu_L$, provided one corrects for setup specific offsets using e.g. a control sample with known values. Figures 3C and 3D show $\mu_L$ and $V$ for >99% pure cyclohexane relative to that of injection grade water measured with a TFP, VIPA, TRBS and SBS spectrometer. In each case a correction factor was applied (see Supplementary Text) based on how the values of $\mu_L$ and $V$ for injection grade water differed from those of deconvolved measurements of injection grade water measured using a TFP in Vienna and Hannover and interpolated for all temperatures (made available at DOI:10.6084/m9.figshare.27794913). The differences in the TRBS results may be due to the high acoustic attenuation of cyclohexane compromising the number of discernible oscillations in the time traces, and thereby the Fourier transformed frequency spectrum.

## Conclusions & Discussion

BLS microscopy is an emergent biological imaging technique that provides a unique contrast based on the high frequency acoustic properties of a material. These can be used to calculate the associated viscoelastic parameters and render near optical diffraction limited spatial maps of these. Being label-free and minimally invasive it shares a comparable platform to imaging modalities such as Raman microscopy. Due to its narrow spectral window of excitation and emission, BLS can be performed simultaneously with other imaging techniques (Raman[73], Fluorescence[74], etc.) allowing for multimodal imaging.

There are a number of differences between the current consensus statement and consensus statements for other spectroscopic techniques[75] and techniques that explicitly probe viscoelastic properties. Most notably, many of the instruments addressed here are largely self-built, and differ both conceptually (time-domain, spontaneous, stimulated) and with respect to their unique sources of uncertainty. These complications are contrasted by there usually only being two key parameters over a narrow frequency bandwidth that are of interest. While good agreement can be obtained between laboratories and instrument designs for the BLS frequency shift derived parameters ($V$, $M'$), there is significant variability in the reported BLS linewidth derived parameters ($M''$, $\eta_L$, $\mu_L$) which are much more sensitive to experimental configuration. Nevertheless, correcting to a standard sample (such as the Vienna-Hannover measurements on injection grade water done here) can bring these into quantitative agreement amongst different laboratories and spectrometer designs.

Currently the broader implementation of BLS microscopy is, to a large extent, limited by the lack of commercial instruments that can be seamlessly integrated into biological imaging research laboratories, with setups for the most part found in laboratories focusing on optical microscopy development. It however has the potential to fill an important gap in the toolbox of technologies that can quantitatively measure changes in mechanical properties of biological specimens with high spatio-temporal resolution in 3D. BLS can also offers unique possibilities, inaccessible to other techniques that probe mechanical properties, such as quantitatively measuring[7,8] and mapping[15,18,19,76] the mechanical anisotropy (different stiffness tensor components), refractive index[24], as well as combined shear and longitudinal viscoelastic parameters[18-21].

Already in the life-science research and pre-clinical areas, the value of establishing a consensus is important to reliably compare measurements performed by different instruments, as well as for the commercialization of spectrometers. For the latter it can serve as a blueprint for parameters that need to be documented, and a guide in designing instruments with performance parameters suited for diverse bio-applications. It may also serve as a springboard for formulating more rigorous and application-specific measurement and reporting standards, needed in diverse commercial and medical applications.

To this end we also advocate for the establishment of a common file format to store raw as well as processed BLS spectral data, together with important metadata to understand the context of any given experiment. For this, the popular and well-accepted HDF5 file format[77] could provide a versatile solution, as it allows one to store complex imagery as well as underlying raw data in a hierarchical and well-structured manner. A first proposed specification for this file format, together with code that allows one to export data to this format, are available in Ref[78]. We foresee to fine tune this format in the future through input from the academic as well as industry community. We expect this standardization in file format to significantly facilitate collaborations, mutual comparisons of experimental results, as well as promote more universal data analysis software development.

While BLS is still in its infancy in regard to clinical translation, one area where it has already made the jump to clinical applications is that of ophthalmology. Here it is being used to identify the severity of pathologies such as keratoconus which are associated with spatial changes in corneal biomechanics[79-81]. As BLS technological sophistication continues to develop in areas such as acquisition speed enhancement[82-85] and reduction of instrumentation footprint[86], it simultaneously is also developing in application specific directions (such as endoscopes[87,88] and versatile fiber-coupled probes[89-92]). One can foresee diverse potential clinical applications that offer improvements over current clinical mechanical sensing (such as palpation and ultrasonic elastography in terms of resolution), as well as novel in-vitro mechano-histopathological[84,93-95], cellular[96,97] and biofluid-based diagnostic applications[90,98-100]. Along such clinical translational pathways additional sample-type specific consensus points will no doubt need to be established, as has been the case for more mature technologies such as optical coherence tomography, photoacoustic imaging, Raman and infrared spectroscopy.

As a field, BLS microscopy can be expected to progress in three complementary directions: (1) where BLS-measured viscoelastic parameters provide unique insight into biophysics inaccessible to other techniques that give fundamental insight into material properties and processes in biological systems. (2) Where empirical or theoretically substantiated correlations with viscoelastic parameters measured at different frequencies and boundary conditions can be made, and it serves as a reliable proxy for these. (3) Where BLS measured parameters provide a unique and useful contrast agent on account of their empirical correlation to a biologically relevant material state or process of interest (e.g. a specific pathology). Here they might in of themselves be accepted as presenting a complex chemo-physical property, related to mechanical properties in so far as the measured hypersonic acoustic quantities are. Regardless of the field's trajectory, it currently finds itself in the somewhat serendipitous position of the technology being ahead of our ability to appreciate the measured parameters' full biophysical/biological significance. To this end establishing consensus, such as the one presented here, early on is particularly important to assure meaningful and consistent progress on all fronts.

## Acknowledgements:

PB and KE acknowledge funding from the Austrian Science Fund—FWF (P34783) and the Medical University of Vienna. CB, JL and RP acknowledge funding from the EMBL, an ERC Consolidator Grant (864027, Brillouin4Life) and the German Center for Lung Research (DZL). SC, DF and MM acknowledge funding from the European Union - NextGenerationEU-Mission 4-CUP B53D2302864 0001 grant. JC, NK and DK acknowledge funding from the DFG (Cz55/44). DH and TL acknowledge funding from the DFG Cluster of Excellence PhoenixD (EXC 2122, 390833453). H.J-R acknowledges funding from the Canada Research Chairs program (CRC-2021-00456). IK acknowledges funding from the ARC Centre of Excellence in Qunatum Technology (CE230100021) and ARC Centre of Excellence in Optical Microcombs for Breakthrough Science (CE230100006). KK acknowledges funding from the National Science Foundation (DMR-2202472). SLC acknowledges funding from the Royal Academy of Engineering Research Fellowships scheme (RF-2324-23-223) and Nottingham Research Fellowship scheme. JM acknowledges funding from an EIC-2022-PathfinderOpen grant (VIRUSong, 101099058). MM acknowledges funding from Research Ireland (13/RC/2073_P2) and Research Ireland Frontiers for the Future Project (22/FFP-P/11394). DRO acknowledges funding from the National Institutes of Health (EY022359). EP and CT acknowledge funding from an EIC-2022-PathfinderOpen grant (ivBM-4PAP, 101098989). GR and CT acknowledge funding from an ERC-2019-Synergy Grant (ASTRA, 855923). GS acknowledges funding from the National Science Foundation (DBI-1942003) and the National Institutes of Health (R21CA258008, R01EY028666, R01EY030063, R01EY032537, R21EY035483). PT acknowledges funding from the Nanyang Technological University, the Ministry of Education (MOE) and the National Research Foundation (NRF). VY acknowedges funding from The Air Force Office of Scientific Research 1189 (FA9550-20-1-0366, FA9550-20-1-0367, FA9550-23-1-0599), the National Institutes of Health (R01GM127696, R01GM152633, R21GM142107, 1R21CA269099) and support from NASA, BARDA, NIH, and USFDA, under Contract/Agreement No. 80ARC023CA002. SH-Y acknowedges funding from the National Institutes of Health (R01EY033356, R01EY034857). JZ acknowledges funding from the National Science Foundation (CBET-2339278) and National Institutes of Health (K25HD097288, R21HD112663).

*Supplementary Text*

# Consensus Statement on Brillouin Light Scattering Microscopy of Biological Materials

Content:

# 1: Different BLS instruments & reporting recommendations

## Spontaneous BLS

Spontaneous BLS, broadly refers to the case where the probed phonons are not being generated/stimulated by some additional means (e.g. by additional lasers or mechanical transducers). To this end there is typically only a single probing laser, from which the scattered light is analysed, to reveal the scattering from inherent thermal phonons. Currently spontaneous BLS techniques are exclusively confined to the frequency domain, namely they measure the frequency spectrum of the scattered light. In the following subsections, we describe some details and specifics to instruments used in spontaneous BLS measurements.

*Single etalon spectrometers*

The simplest setup for a Brillouin is a non-scanning single pass, single etalon system. This yields a standard Fabry-Perot set of fringes where the center is the zeroth order and each successive mode is a higher order with non-linear dispersion between orders. Orders are Rayleigh peaks; surrounding each order is the set of Brillouin Stokes and anti-Stokes pairs. In a single etalon system, the light has to be angularly dispersed through the etalon either by tilting or diverging the beam. A cylindrical lens can be used to compress the dispersing light to a line rather than illuminating the entire circular Fabry-Perot pattern. FSR range between orders requires calibration and linearization yet multiple orders can be used for spectroscopic information. Experimental parameters important to report with a single etalon system are the manufacturer, air-gapped (and the gas if different than air) or solid etalon, the flatness, free-spectral range, finesse, how many orders are used in analysis, and whether the spectra is linearized or not linearized.

**Multi-pass Fabry-Pérot interferometer**

A Tandem Fabry-Pérot spectrometer (TFP) relies on a tandem combination of two air spaced etalons whose spacing is precisely adjusted and controlled. By varying the mirror separation of the etalons, the device sequentially scans consecutive wavelengths. Multiple passes through the etalons are used to increase contrast and finesse. One popular example of this is the 6-pass TFP spectrometer TFP-2 HC, a commercial device developed by Table Stable Ltd., which can achieve a contrast of up to $10^{15}$. This high contrast can suppress the elastic background in even strongly turbid samples[73].

The FSR can be varied between about 5 THz and 5 GHz, with a finesse of 100-120 giving a maximum resolution of about 50 MHz. The standard scan time is ~500 ms, but the speed and acquisition bin size can be modulated allowing for e.g. selected spectral windows to accumulate more photons quicker and thus for faster measurements. Using external timing sources, a commercial TFP-2 HC allows faster scanning up to a factor 3-5. A faster version of the TFP, capable of reaching full scan times of 30 ms without any loss of performance, has been developed successfully by the company Table Stable Ltd. and will be commercialised soon.

A major advantage of the scanning Fabry-Perot is that it is an absolute instrument: the wavelength calibration is determined uniquely by the mirror spacing which is typically known to ~0.1% or better. As such no calibration standard is required, and under laboratory conditions such an instrument will stay stable for hours and even days. Transmission efficiency can be checked by comparison to known samples. When frequency locked to the laser wavelength, there is no theoretical limit on how long such an instrument can be held stable.

When publishing measurements, in addition to the spectrometer manufacturer, one should report the mirror spacing, pinhole sizes (entrance pinhole to spectrometer and exit pinhole to detector), scan amplitude(s), total recording time per channel, Quantum efficiency (QE) of the photon detector used, and a full description of the optics feeding the signal into the TFP.

**VIPA-based spectrometers**

VIPA (Virtual Imaged Phase Array) spectrometers are non-scanning systems that capture the entire BLS spectrum simultaneously. A VIPA is a type of Fabry-Perot etalon in which the upstream partial reflector is replaced by a high reflector, except for a small window through which the light is coupled. The dispersion of a VIPA is the same as that of a standard angularly-dispersed etalon. While the latter has a throughput on the order of inverse finesse, the transmission of a VIPA approaches unity for a near diffraction-limited input along the dispersion axis (e.g. confocal collection or line scan geometry)[101]. Consequently, depending on the excitation conditions and sample, VIPA spectrometers can acquire a Brillouin spectrum in a few tens of milliseconds. A FWHM resolution of a few 100 MHz and a sub-10 MHz precision are typical. The VIPA thickness can be determined within 0.1%, facilitating accurate absolute calibration. The simplest VIPA spectrometer setup consists of a collimated beam focused with a cylindrical lens into the entrance window of a VIPA, followed by an imaging lens and a camera to capture the spectrum. The contrast of a single VIPA being inherently limited to ~30 dB, several cross-dispersion configurations and elastic peak suppression strategies have been developed as described below.

**Crossed VIPAs:** To increase the contrast of spectrally-dispersive optical elements, cross-axis cascading of multiple elements[102] offers an efficient option that has been used in many VIPA-based spectrometers. The principle of cross-axis cascading with VIPA etalons is straightforward: if a first stage of optical dispersion is aligned along a vertical direction, then the spectral pattern is also dispersed vertically. In this situation, any elastic scattering component beyond the contrast of the first spectral stage appears as a crosstalk signal along the vertical spectral axis. If a second stage of spectral dispersion is now aligned orthogonally to the first stage, the spectral pattern exiting the first stage enters the second etalon through the input window. Since both etalons disperse light in orthogonal directions, the overall spectral axis of the double-spectrometer lies along a diagonal direction, which is where the Brillouin spectrum appears. Instead, the crosstalk elastic signal, which was spatially overlapped after the first stage, after the second stage, is mostly confined to the horizontal and vertical axis. This procedure has been shown to be repeatable for multiple stages[103], but the double-stage configuration has proven to provide the best tradeoff between throughput and contrast. In most realizations, the elastic scattered light is physically blocked with an optical mask in an intermediate image plane or planes to avoid detector saturation. The crossed-VIPA architecture is compatible and often used with other improvements that have been demonstrated over the years such as apodization[104], coronography (Lyott stop)[105] or diffraction masks[106], leading to high-throughput contrast of ~80 dB. These combinations usually suffice for measuring e.g. isolated biological cells, but not for more opaque samples.

**Crossed VIPA-echelle grating:** Alternatively to crossed VIPAs, contrast enhancement through cross-dispersion can also be achieved with an echelle grating. The latter displaces the Brillouin signals away from the VIPA Lorentzian tails of the typically much stronger elastic signal. This results in a contrast on the order of 50 dB at the water BLS peak position and up to 60 dB beyond. The grating also separates otherwise overlapping VIPA orders, producing an unambiguous simultaneous range of several THz with minimal spectral artifacts. This enables the BLS and ultra-low frequency Raman measurement of a large range of samples in a snapshot. Since this configuration produces a 2D pattern on the camera, calibration involves mapping which pixels can be illuminated and determining to what wavelength they correspond based on the VIPA and grating dispersion equations. Commercial solutions are available employing such a crossed VIPA-echelle grating scheme, which when combined with an elastic suppression filter can achieve an effective contrast of 120 dB[107]. Alignment of the crossed VIPA-echelle grating scheme is similar to a single VIPA, since the grating is essentially alignment-free.

When publishing measurements on VIPA-based spectrometers additional parameters that should be reported are the VIPA FSR, the spectral interval measured on the detector array, the number of orders used to analyze the BLS peaks, and if relevant the model number in the case of a commercial instrument.

**Elastic suppression filters:** An alternative solution to cross-dispersion that has been used to gain visibility of the BLS spectral peaks, in primarily VIPA-based spectrometers, is through the employment of special high-rejection narrow-band optical filters to attenuate the elastic scattering light. In this case, the elastic background light is suppressed by a notch filter, relaxing the need for multiple optical passes and/or cascaded dispersive elements to achieve high spectral contrast at detection. Employing a sufficiently effective filter can allow one to employ only a single VIPA spectrometer to perform measurements on turbid media.

Several methods have been demonstrated to efficiently suppress the elastic background light with extinction ratio ranging from 30 to 50 dB. Commonly employed methods include the use of etalons[108,109], absorption cells[110] or interferometric schemes[111,112]. Two popular filter types currently being employed are:

*Gas absorption cell filters:* Atomic and molecular line absorption filters can provide an efficient approach to suppress the elastic scattering in BLS microscopy. Since some of the most commonly used laser wavelengths for BLS microscopy (532 nm and 780 nm) are close to molecular absorption lines of molecular Iodine (532 nm) and atomic Rubidium (780 nm), the use of absorption lines from these to suppress elastic scattering can be used. Here the incident laser light can be tuned match a selected absorption line and be actively stabilized and frequency locked to this line[113]. The advantages are the relatively low cost and easy maintenance of such a filter, which are independent of environmental conditions and beam divergence. The Doppler and collisional linewidth are insignificant for most practical conditions, and a high degree of attenuation can be achieved using multipass geometry or longer cell pathlengths[35,114].

Potential drawbacks of gas absorption cells include the presence of a number of absorption lines that may modify the transmission spectrum and thus final measured spectra (particularly the case in Iodine cells), residual emission from the cell, and a potentially large variation of transmission for a cell whose temperature is not actively stabilized. Some of these issues can be addressed through BLS measurements taken at multiple excitation wavelengths, at the expense of making the overall setup more complicated and less suitable for rapid acquisition[115]. Gas absorption cells also ideally require the use of laser lock-in scheme, which typically comes with an increased optical system complexity. When using gas absorption cells, it is recommended to report the chemical composition of the cell and its dimensions, the number of passes through the cell, the manufacturer and model of the cell, and the temperature of the cell when in use.

*Etalon-based filters:* A low-insertion loss, ultra-narrow notch filter can be achieved with a Fabry-Perot etalon tuned (angle, temperature, pressure, or piezo) to transmit the excitation wavelength while reflecting the Brillouin signals. The transmitted excitation wavelength light is then discarded, while the light covering the BLS peaks is coupled into the BLS spectrometer, typically via a fibre or small pinhole for spatial filtering. The alignment consists of optimally centering the filter transmission at the laser wavelength, and subsequently optimising the coupling into the output fibre or pinhole. Commercial solutions, by the company Light Machinery, based on a 4-pass pressure-tuned etalon which produces up to 70 dB suppression are available. Other types of interferometers can be used to filter out the elastically scattered light, such as modified Michelson interferometers[112] or prism-based interferometric filters[116].

While the development of integrated filters is ongoing[117], a common-path birefringence-induced phase delay (BIPD) filter providing up to 60 dB extinction ratio in a single pass has been recently proposed[65], and made commercially available by the company *Specto Srl*. This BIPD filter can be optimized for all visible wavelengths without the need of sophisticated laser lock-in systems.

In general, when using such elastic filter(s) one should report their FSR, the type and arrangement of the filter(s), their experimentally obtained extinction, and the manufacturer(s) and model(s) of any elements employed in their realization.

**Radial VIPA:** Recently the concept of the VIPA has been extended by replacing the entrance *slit* of the VIPA with a *hole*, effectively changing the Cartersian degree of freedom the conventional VIPA offers (which is exploited in e.g. Line-scanning BLS[82,83]) to a radial one[15]. With a suitable excitation scheme this allows one to instantaneously measure (image) the frequency shift, and thereby acoustic speed and longitudinal modulus, in different (azimuthal) directions in a sample and determine the mechanical anisotropy. Being by construction limited to a single pass/VIPA configuration, it requires good suppression of the elastic scattering (as may be achieved by e.g. absorption cells—see above), and currently still only suited to reasonably transparent samples. For dynamic measurements and measurements with low (physiologically acceptable) laser exposure it is necessary to perform angular binning to obtain statistically useful spectra. In addition to reporting the typical parameters for VIPA spectroscopy such as the FSR, and that of elastic scattering suppression schemes described above, it is also important to report the range of wavevectors/scattering-angles probed in the sample (determined by the optical setup before the spectrometer) and the azimuthal and polar angles that are being averaged over for each spectral projection.

## Stimulated BLS

Stimulated Brillouin scattering (SBS) is a nonlinear phenomenon driven by the interaction between optical and acoustic fields through electrostrictive, absorptive, and photoelastic processes under energy and momentum conservation conditions. In SBS, induced density variations (i.e., an acoustic wave) in the medium are typically driven by the interference of two incident optical waves which are detuned in temporal frequency and/or spatial frequency around the BLS resonances of the medium. The stimulated acoustic wave can then be observed in the backward direction as an increase (or a decrease) in the intensity of a weak optical wave or can be interrogated in the forward direction by a third readout wave[118]. There are currently two types of BLS microscopy that are referred to as SBS microscopy (*Frequency domain* SBS and *Impulsive* SBS), although Time-Resolved Brillouin Scattering (TRBS) also technically falls into this category (and is thus included here).

SBS microscopy's major limitation is the complexity of the experimental arrangement and the difficulty of the alignment of counterpropagating beams and scanning one of the frequencies. In its earliest implementation[119], the primary challenge came from the signal strength limited by the allowed pump power; however, through pulsed excitation[120,121] and quantum-enhanced imaging[122] the problem of high average power lasers can be largely mitigated. Alternatively the laser exposure can be reduced significantly by only measuring at discrete frequencies[84].

*Frequency domain SBS*

In frequency domain SBS, two counter-propagating, frequency-detuned optical fields interfere in the medium[26,123]. One field is relatively intense (pump) and the second is typically weaker (probe). The wave interference induces travelling density variations (i.e. an acoustic wave) in the medium primarily via electrostriction. This consecutively produces a travelling refractive index variation through photoelasticity. Under proper incidence polarisation circumstances and frequency detuning that closely matches Brillouin lines of the medium (on the order of 5 GHz at near infrared wavelengths), the pump intensity is reflected from the moving grating with a Doppler shift of the probe light. As a result, the intensity of the probe increases (decreases) with a fractional gain (or loss). The corresponding lifetime of the acoustic wave is ~1 ns at near infrared wavelengths, resulting in BLS linewidths of several hundreds of MHz. When publishing SBS results it is additionally important to report the frequency range over which the probe laser is scanned and the corresponding frequency steps.

*Impulsive SBS*

In Impulsive Stimulated Brillouin Scattering (I-SBS), two pulses interfere to stimulate acoustic phonons within the sample. The readout process involves Bragg diffraction of a probe beam, commonly a CW laser at a different wavelength, which results in an intensity-modulated time-signal[85,124]. Unlike above mentioned techniques, I-SBS measures in the time domain, allowing for potentially higher time resolution, albeit at reduced spatial resolution. The excited acoustic frequency ($f$) in I-SBS can be calculated via $f = 2V/d$ f = 2*v / d (for electrostriction), where $V$ is the sound speed, and $d$ is the interference fringe spacing. The frequency is tunable by adjusting the intersection angle $\theta$ of the pulse beams. The frequency resolution of I-SBS, or the ability to distinguish between two frequencies in the spectrum, is inversely proportional to the signal length.

The signal length depends predominantly on the size of the excitation volume. A high pulse energy is required to generate high SNR, but requires a good balancing with pulse length and repetition rate. When publishing SBS results it is thus additionally important to report laser pulse characteristics. The interference fringe spacing of the two excitation beams acts as a lower limit for the spatial resolution, because the probe beam has to diffract at least one acoustic period. One possible application of I-SBS can be *Brillouin cell cytometry*[125] where a typical spot size of 10 μm is employed.

*Time resolved BLS*

Time-domain Brillouin Light Scattering (TRBS), also referred to as *Phonon Microscopy* in the literature, is a collection of techniques that are generally characterised by the following two conditions: (a) there is a photoacoustic generation of coherent acoustic phonons using a pulsed pump laser and a transducer (typically a thin metallic film). (b) One measures the temporal modulation of the intensity of a second laser beam probing the propagation of the generated phonon, that is representative of the phonon time-of-flight[64,126].

Compared with other BLS imaging techniques, time-domain is capable of spatial resolutions down to the acoustic wavelength, and also provides access to the instantaneous phase of the acoustic wavefront. To construct a time-domain BLS optical system, it is important to consider the 3D spatial overlap of the pump and probe beams. For the initial stages of alignment, it is most convenient to use the top surface of the opto-acoustic transducer as the reference plane. The pump laser should be collimated entering the back aperture of the objective lens. Due to the incompatibility of shearing interferometers with ultrafast pulsed lasers, collimation is verified by visually inspecting the beam size at infinity while adjusting the length of the beam expanding lens pair. The position of the brightfield camera can then be optimised such that the sample and pump beam are both in focus. Next the process is repeated for the probe beam to match the focal plane of the pump and brightfield optics.

The transient reflectivity or transmissivity of the probe beam is typically detected by a ~5.5 MHz bandwidth amplified (10dB) photodiode, further amplified (24dB), and low-pass filtered (11 MHz) before digitisation by an oscilloscope. The time signal on the oscilloscope should be maximised in amplitude by adjusting the x, y, and z overlap of the pump and probe beams. Once maximised the signal can be processed according to the protocols set out in[92] to extract the phonon time-of-flight, which should be normalised by the DC light level (equivalent to the amplitude of the Rayleigh peak) to achieve units of modulation depth. This processed modulation depth signal can be analysed in the frequency domain through Fast Fourier Transform to measure the BLS frequency shift $\nu_B$. The phonon attenuation rate is related to the longitudinal viscosity and can be measured directly in the time-domain by fitting a decaying exponential to the TRBS signal. The decay constant ($\alpha$) is typically interpreted in units of µm$^{-1}$ by converting the time-base of the signal into the axial spatial domain through the inferred relationship between the BLS frequency shift and sound velocity. However, in order to provide an equivalent BLS linewidth measurement in the frequency domain (after taking an FFT of the time-signal), the finite nature of the time window must be considered as this introduces a spectral artifact in the form of the Fourier transform of a Heaviside function (truncation of the time window). This can be calculated analytically and results in the following correction formula: $\Gamma_B \approx \Gamma_{TRBS} - 3.78/(2\pi T)$ where T is the length of the time window and $\Gamma_{TRBS}$ is the full width half-maximum of the Brillouin spectral peak.

It is recommended to report the modulation depth and SNR of a given time-domain BLS setup, along with experimental parameters such as the average optical power, NA of the illuminating objective lens, laser repetition rate, and pulse width. The modulation depth scales with the acoustic amplitude (proportional to pump intensity) and probe intensity: I$_{pump}$I$_{probe}$. Consequently, the specimen type will dictate the sustainable pump and probe intensities (and resulting transducer heat rise). Utilising sapphire substrates and pump and probe powers of 1 and 3 mW respectively, a modulation depth of ~10$^{-4}$ and SNR of ~70 can be obtained[64] using a transducer with 20:160:20 nm Au:ITO:Au layers. For living cells less power should be used (0.5mW pump and 1mW probe) and typically modulation depths on the order of 10$^{-6}$ and SNR of~40 are obtained with transducer heating on the order of ~1°C[64,127]. To estimate the SNR of a time-domain signal, the amplitude of the BLS peak is divided by the standard deviation of amplitudes in the band of interest from an additional measurement with the pump beam blocked. The SNR will scale proportionally to the pump and probe intensities, SNR ∝ $I_{pump}\sqrt{I_{probe}}$ with typical values in the range of 40-70.

Accurate measurement of the BLS frequency shift from time-resolved Brillouin scattering (TRBS) signals can be affected by artefacts generated from several signal components. When both pump and probe pulses overlap in time, optical absorption induces a strong electronic excitation within the metallic transducer and results in a sharp signal response known as the coincidence peak. This manifests itself as a rapid change in the optical properties, temperature, and ultimately the propagation of coherent phonons. The response typically relaxes according to the thermal diffusivity of the system before the arrival of the next coincidence peak. Once the sample volume has been excited mechanically there are additional frequency components in the signal which should be digitally filtered out. The mechanical resonance of the transducer (~10GHz) in sapphire decays after only one or two cycles but it can be more prominent on softer substrates; and the BLS frequency shift of the substrate material which for stiff materials exceeds the detection bandwidth of the system can be digitally low-pass filtered.

To isolate the TRBS signal from these components, the following digital operations may be performed in post-processing[128]: cropping the signal after the coincidence peak, low-pass filtering, and subtraction of the thermal background via low-order polynomial fitting. Cropping too close to the coincidence peak can cause reduced SNR and BLS frequency shift measurement precision due to the presence of broadband signal components, and cropping too far artificially attenuates the desired signal. Similarly, a polynomial fit order that is too low can cause errors in the determination of the BLS frequency shift, whereas an aggressive polynomial can act as a high pass filter.

The frequency resolution of the system is given by the repetition rate of the lasers (80-100MHz) which is typically insufficient to detect small variations in the BLS frequency shift in biological tissue. However, since the bandwidth of the signal is much greater than the frequency bin spacing, the spectrum can be interpolated by zero padding before calculating a Fourier transform ($2^{14}$-$2^{16}$), allowing one to determine the centre frequency of the spectrum with ~10 MHz precision[92]. In the case of estimating the longitudinal sound attenuation coefficient, it is important to note that a short optical depth of focus at high NA can lead to measurement errors. This is due to the signal decaying by loss of optical intensity rather than acoustic intensity. The acoustic attenuation in water at 5.1 GHz and at room temperature is ~$2.5 \times 10^5$ m$^{-1}$, and comparing deviations to this value can provide a means for correction[29].

In time-domain BLS the photo acoustically-stimulated coherent phonon field occupies a well-defined region in space, defined laterally by the PSF of the pump laser at the transducer interface, and axially by an exponentially decaying envelope set by the phonon path length, and with a well-defined wavevector (approximately a plane wave). It follows that the fundamental limit for lateral resolution is defined by the convolution of the spatially overlapped pump and probe PSF. Since the measured time of flight of the scattered probe beam is proportional to the axial distance, its period is that of the acoustic wavelength (which is shorter than the optical wavelength by a factor $\lambda_{optical}/2n$), and time-frequency analysis allows for measurements of $\nu_B$ with an axial resolution down to that of the acoustic wavelength[91,127]. Conceivably the lateral spatial resolution can also be reduced to that of the phonon wavelength if one were to employ opto-acoustic lenses and time of flight analyses.

Some parameters important to additionally report in TRBS are the modulation depth and SNR of the TBBS setup, as well as the repetition rate and pulse width of all employed lasers.

## Heterodyne BLS

Heterodyne detection is a potent technique utilised in various optical measurements, including BLS. In standard BLS detection, the measured signal is a measure of the number of photons collected within a given bandwidth and integration time, that is proportional to the square of the scattered electric field. Heterodyne detection entails mixing the scattered light with a reference beam, termed the local oscillator (LO), at a slightly different frequency. The interference between the two beams generates a beat frequency in the radiofrequency (RF) range, which is linear with the weak BLS electric field and the strong LO field. This setup often operates with LO power around 1 mW and scattered light power in the femto-Watt range, shifting the challenge from low light optics to low electric signals in the sub-GHz region.

An analog heterodyne detection setup[129] typically includes several key components. A monochromatic light source passes through the sample, where it undergoes BLS. The scattered light is then mixed on a square law photo-detector with the LO beam. A microwave driven electro-optic amplitude modulator generates the LO by shifting the source at specific frequencies close to the Brillouin peak. This brings the beat frequency into the sub GHz range, where the two electric signals are amplified, filtered using a bandpass filter, and further cleaned up by a lock-in based detection. The width of the bandpass filter determines the spectral resolution of the BLS measurement.

A digital heterodyne detection setup[130], on the other hand, incorporates high-speed digitizers and Field-Programmable Gate Arrays (FPGA) for data processing. Similar to the analog setup, the scattered light and EOM generated LO beam are mixed, producing the beat frequency. The digitizer samples the beat signal at high rates, such as 3.2 Gs/s, and the FPGA processes the data in real-time. This digital approach offers higher processing speed and flexibility, enabling real-time data analysis.

Heterodyne detection is an intrinsically single mode measurement, thus the usual single mode fibre collection of Brillouin microscopy confocal layout is best suited for this approach.

The strength of this approach is that, thanks to the down conversion of the optical heterodyne mixing, contrast and resolution are no longer a major issue, as they are not limited by the transfer function of optical devices, but benefit from the tools and precision offered by electronics in the sub GHz region. As a consequence the approach promises a calibration free, extremely stable and compact layout.

When using a heterodyne detection setup it is additionally important to report the manufacturer(s) and model(s) of instruments used, the achieved spectral resolution , as well as characteristics of the bandpass filter and when relevant those of the Analog to Digital Converter (ADC).

**Fibre probes**

BLS fibre probes are devices made of or compatible with optical fibre components that can replace the confocal microscope ("front-end") of a BLS imaging system. Different designs have been proposed and demonstrated to date based on whether signal acquisition occurs in the frequency or time domains.

For the frequency domain, BLS fibre probes can be constructed using a single or dual optical fibre configuration and then supplemented by a standard VIPA or a scanning TFP spectrometer to produce a fibre-integrated BLS measurement setup. The main drawback in using optical fibres for delivery of the laser light and for the collection of the backscattered light, is the spontaneous BLS that occurs inside the fibre core. Since light is tightly focused inside the core of the optical fibre, and a typical length of an optical fibre is a few metres, the effective interaction volume between light and fibre core material is significantly greater than the volume of the optical voxel formed by the focussed beam in the sample. Thus, despite the fibre BLS frequency shift usually being significantly larger than that of the probed sample (>20 GHz), this parasitic signal might present a challenge in detecting the BLS signal originating from the sample[35]. Different solutions to this problem exist, including: 1) hollow-core optical fibres for delivery of the laser illumination to the sample without inducing any significant Brillouin scattering in the fibre itself[88]; 2) the use of dual-core or dual-fibre geometries, as is done in Raman endoscopy, where delivery and collection channels are separated thus avoiding mixing of the fibre backscattering signal with that from the sample[87]. Dual fibre approach requires careful design of the focusing optics in order to ensure collection of the inelastically scattered light back into the core of the second fibre. This can be done by creating on demand 3 dimensional microoptics and/or compound lenses fabricated via two-photon-polymerisation laser writing technique. Overall, the imaging resolution in lateral and axial directions depend on the focusing micro-optics design and is typically slightly worse when compared to conventional microscopy objective lenses, owing to low numerical aperture of the single-mode optical fibre and fibre-compatible lenses[88].

Brillouin fibre probes can also be constructed to operate on the principles of TRBS; however this requires fabrication of an optoacoustic transducer onto the fibre tip for the generation of coherent phonons. These stimulated phonons provide an enhanced BLS cross-section, however currently at the cost of slower detection electronics. The time-domain BLS process is largely insensitive to BLS back scattering along the glass fibre - since the high amplitude stimulated phonons are localised to within 10-20 µm of the transducer as is the coherence length of the interaction. This means that standard silica fibres of arbitrary length can be used.

A time-domain BLS system can be converted for fibre implementation by fibre coupling the free-space pump and probe beams and coupling these into a common channel via a 2:1 fibre coupler. The common channel contains an inline fibre circulator, containing the sample-facing fibre at one port and the detection photodiode at the other, thus avoiding losses incurred by using standard 3dB couplers. If the pump and probe beams contain separate wavelengths, then SNR can be further improved by using a wavelength division multiplexer to attenuate the pump beam at the detector and reduce its contribution to shot noise[92]. SNR and modulation depths will be comparable to free-space TRBS provided similar parameters are used for the optical system and optoacoustic transducer.

Time-domain fibre probes have been shown can achieve 2 µm lateral resolution, realised by point-scanning the sample or fibre, and down to 260 nm depth resolution (equivalent to the phonon wavelength) which is achieved without confocal scanning the optics since the time-domain signal encodes the depth domain[91]. However, high depth resolution comes at the cost of reduced depth measurement range (~10 µm) which can be increased by lengthening the probe wavelength (the attainable depth scales as the square of the probe wavelength).

The exact details of the fibre inelastic scattering depend on the type of fibre used, its length, the amount of bending of the fibre, and other properties of the experimental configuration. Even the use of hollow-core optical fibres does not remove the fibre background completely since a small percentage of the fundamental fibre mode and higher order modes interact with the microstructured region of the hollow-core fibre, producing a small amount of backwards-scattering BLS[88]. Such a parasitic signal, however, does not interfere with the desired BLS signal from the biological sample, as it is positioned outside the spectral region of interest, and can be removed during data post-processing. High order fibre modes may also be attenuated by bending the hollow-core fibre slightly, without inducing significant bending losses to the fundamental fibre mode.

As with spontaneous BLS fibre probes, in TRBS fibre probes, the BLS signal from the glass fibre is typically low-pass filtered in post-processing. It is worth noting that the time-domain BLS process is only sensitive to the stimulated coherent phonons within the vicinity of the transducer, and although these are approximately 6 orders of magnitude greater amplitude than the spontaneous thermal phonons along the fibre length, they attenuate very rapidly, travelling away from the transducer within tens of microns. This allows one to use long lengths of glass fibre without adversely affecting the glass signal amplitude. Additionally, the acoustic impedance and photoelastic coefficient mismatches between the glass, metal and water layers result in a ~2 orders of magnitude greater signal amplitude from the water or cell tissue interface compared to that from the glass interface. Sample heating should be considered when performing fibre-based time-domain BLS both in terms of specimen viability and changing the measured BLS frequency shift. The former can be mitigated through lowering optical fluence and the inclusion of thermally conductive transducer layers. Any thermal gradient produced by the fibre probe can be calibrated by obtaining reference measurements (away from a specimen) in the aqueous couplant medium. These have been found to be stable under normal experimental conditions, such that subtracting these frequency offsets from the specimen BLS shifts can potentially compensate for such thermal contributions[91,92].

When using fibre probes it is additionally important to report the manufacturer and model of the optical fibre, any strategies used to overcome the contribution from BLS backscattered light from the fibre, the coupling efficiency of reflected and/or Brillouin scattered light back into the fibre-core prior to detection, and as relevant any treatment performed on the tip of the fibre. When using transducers at the end of fibre probes, it is additionally desirable to present an assessment of the sample heating induced by the transducer, and any relevant calibrations made for this.

## 2: Some useful equations for BLS microscopy

The interpretation of Brillouin Light Scattering (BLS) measured parameters is historically divided between formalisms that consider the measured parameters in two different frameworks.

Firstly, there is that of acoustic (e.g. shock) waves moving through a solid and largely elastic material, which is relevant for describing e.g. geological activity, mechanical properties of polycrystalline materials, etc.. Here the material properties are often treated as effective entities, with the interest typically being how the collective, and for the most part elastic, properties change e.g. in different directions. Given the characteristic relaxation times of *hard* solids are very slow, the BLS derived *stiffness coefficients* can be useful for modeling also lower frequency acoustic and mechanical properties.

Secondly, there is the thermodynamic and (molecular) hydrodynamic picture, which considers the molecular interactions on short time scales to predict the supported phonon modes. From this one can gain insight into the electrostatic screening of polymers (such as DNA) and structural states of molecules that affect their interactions with a solvent or molecules in their immediate vicinity. Here the time scales probed using BLS may become more comparable to those of the relaxation processes, and one can gain insight into the nature of e.g. phase and structural transitions by virtue of the changes in the hypersonic speed and attenuation. The interpretation here is less on our everyday understanding understanding of viscoelastic moduli, but rather on elucidating changes in "viscoelastic properties" on time scales that affect the molecular interactions, namely relaxation times, and how these may lead to changes in both molecular processes and collective material properties.

Biological systems, such as a living cell, rarely fit either of these two, but rather often exist in a somewhat undefined dynamic state between an anisotropic solid and a liquid, usually containing elements of each. In regard to building on the established BLS nomenclature this presents a unique challenge. Below we define some parameters, taken from these two established branches, that can prove useful for analyzing BLS measurements in biological systems. These include both basic fitting parameters and biophysically relevant parameters, with a brief justification of their relevance in each case. These are intended to serve as a potentially useful, but in no way complete, reference for people performing BLS on bio-relevant matter, to assure that all parameters are defined consistently.

**Fitting equations**

*Damped Harmonic Oscillator:* In the frequency domain, the intensity of each BLS peak is functionally described by a Damped Harmonic Oscillator (DHO) given by:

$$I(\nu) \propto \frac{\Gamma_B \nu_B^2}{(\nu^2 - \nu_B^2)^2 + (\Gamma_B \nu_B)^2} \quad \textbf{Eqn. S1}$$

where $\nu$ is the frequency shift relative to the probing laser frequency, and $\nu_B$ and $\Gamma_B$ are the BLS frequency shift and BLS linewidth. If $\nu_B$ falls in the vicinity of a structural relaxation processes ($\sim 1/\tau$, where $\tau$ is the viscous relaxation time) the shape of the peak will generally have a more complex form, and its spectral shape needs to be calculated from first principles[1,22].

*Lorentzian Function:* To an, often good, approximation the BLS peak can also be described by a Lorentzian function:

$$I(\nu) \propto \frac{\Gamma_B^{(L)}}{\left(\nu - \nu_B^{(L)}\right)^2 + \left(\Gamma_B^{(L)}/2\right)^2} \quad \textbf{Eqn. S2}$$

where $\nu_B^{(L)}$ and $\Gamma_B^{(L)}$ are the (Lorentzian) BLS frequency shift and BLS linewidth. If the linewidth is not too large this is usually a good approximation.

*Correction to Lorentzian fit:* Due to the general asymmetry of the DHO, the Lorentzian fitting parameter $\nu_B^{(L)}$ will differ from the true (DHO obtained) $\nu_B$ by an amount that becomes more significant with increasing linewidth given by:

$$\nu_B = \nu_B^{(L)} \sqrt{[1 + (1/2)\Gamma_B^{(L)}/\nu_B^{(L)}]} \quad \textbf{Eqn. S3}$$

**Frequency shift**

The frequency shift $\nu_B$ of a given BLS peak can be expressed in terms of the hypersonic speed $V_q$ of the respective measured phonon modes in the direction of the scattering wavevector $q$:

$$\nu_B = (2\pi)^{-1}\omega_B = \pm(2\pi)^{-1}qV_q \qquad \textbf{Eqn. S4}$$

The scattering wavevector $q$ (which is also the wavevector of the created/anihilated phonons) is given by:

$$q = 4\pi n \sin(\theta/2)/\lambda_0 \qquad \textbf{Eqn. S5}$$

Here $\theta$ is the scattering angle (=180 degrees in the back-scattering geometry), $n$ is the refractive index, and $\lambda_0$ is the free-space wavelength of the probing laser.

In the vast majority of BLS microscopy studies on biological samples (and to date practically all studies on living cells) one measures only the *longitudinal acoustic phonons*. These in almost all cases have larger BLS frequency shifts than the transverse phonons also supported in solids and some liquids, and a much larger scattering cross section. They correspond to travelling density waves where the displacement of molecules is entirely in the direction of the propagating waves. Namely, solutions of the wave equation:

$$\frac{\partial^2 u_i}{\partial t^2} = V^2 \frac{\partial}{\partial x_j}\left(\frac{\partial u_l}{\partial x_k}\right) \qquad \textbf{Eqn. S6}$$

in which $u_i$ is the displacement of molecules in the direction $x_i$, and $t$ is time, with $i = j = k = l$. The speed $V$ here is related to the mass density ($\rho$) of the material, as well as its stiffness. The relevant stiffness here depends on the direction of the strains relative to the stresses. These can be expressed in terms of the generalized Hooke's Law:

$$\sigma_{ij} = c_{ijkl}\varepsilon_{kl} \qquad \textbf{Eqn. S7}$$

Where $\sigma_{ij}$ and $\varepsilon_{kl}$ are the strain and stress tensor (and the subscripts indicate the decomposed orthogonal vector directions, i.e. $i = 1,2,3$). $c_{ijkl}$ is the so-called *stiffness tensor* and describes the complete elastic response of a material in an arbitrary direction subject to a stress in the same or a different arbitrary direction. $c_{ijkl}$ is a huge tensor, but owing to several symmetry constraints only has 21 independent components and is usually more compactly expressed as a 6 x 6 tensor $c_{ij}$. The first three diagonal components ($c_{11}$, $c_{22}$ and $c_{33}$) will correspond to longitudinal modes (i.e. where $i = j = k = l$) in the three orthogonal directions.

In general the relation between the hypersonic speed in the direction of the probed scattering vector $\hat{q} = q/|q|$, and the relevant stiffness tensor components can readily be obtained from the Christoffel equation:

$$\left|(\hat{x}_k \cdot \hat{q})(\hat{x}_l \cdot \hat{q})c_{ijkl} - \rho V_q^2\right| = 0 \qquad \textbf{Eqn. S8}$$

where $\hat{x}_i$ are unit vectors in the direction of the respective subscripts. It follows that the speed of the longitudinal phonons will in the direction $\hat{x}_i$ be given by:

$$V_i = \sqrt{(c_{ii}/\rho)} \quad \text{Eqn. S9}$$

where $c_{ii}$ describes the response to a stress and strain in the same direction ($i$).

The longitudinal elastic modulus will normally correspond to the stiffness tensor components $c_{11}$, $c_{22}$ or $c_{33}$ when the material is measured in a symmetry direction such as along a symmetry axis for a crystal, along the fibre axis for a fibre, or in essentially any direction for an isotropic material like an amorphous material, gel, or liquid. Conceptually the longitudinal elastic modulus can be thought of as the relevant modulus when the strain is uniaxial. This means that the material is stretched or compressed only in a single direction and does not change in the perpendicular directions.

If one assumes that the material is mechanically isotropic (has same elastic modulus in all directions) then $c_{11} = c_{22} = c_{33} = M'$ and it follows from Eqn. S4, S5 and S9 that:

$$\nu_B = 2n \lambda_0^{-1} \sqrt{\frac{M'}{\rho}} \sin\left(\frac{\theta}{2}\right) \quad \text{Eqn. S10}$$

which is the key equation used to calculate the longitudinal elastic modulus from BLS frequency shift measurements. It is important to note that this is for a single scattering angle ($\theta$), whereas one in practice measures over a finite range of scattering angles (on account of the finite NA of the microscope objective), which will result not only in peak broadening but also potential shifts in $\nu_B$. To this end one should ideally integrate Eqn. 1 over the solid probing and collection angles, especially for scattering geometries deviating from back scattering, to obtain accurate measures of $M'$.

The above assumes that not only the longitudinal elastic modulus but also that the refractive index ($n$) is isotropic. The latter may not always be the case, e.g. in fibrillar structures such as muscles or cellulose. In such cases its directional dependence needs to explicitly be accounted for, which may be done by writing it in the form of a tensor ($n \to \boldsymbol{n}$) and considering its projection in the direction of the scattering wavevector. The BLS frequency shift can then be compactly expressed as:

$$\nu_B = \pm 2\, \boldsymbol{n} \cdot \hat{\boldsymbol{q}}\, \lambda_0^{-1}\, V_q \quad \text{Eqn. S11}$$

Where $V_q$ in a mechanically anisotropic sample is in general a combination of stiffness tensor components defined by the projection of the scattering wavevector that can be calculated from Eqn. S8.

It is also possible to measure BLS scattering from transverse acoustic phonons. These show up as peaks at (usually) lower frequency shifts than the longitudinal phonon peaks and typically require the implementation of distinct (non-backscattering) measurement geometries. From these it is possible to calculate the *shear modulus* ($G$) (also called the *transverse modulus*), with the real part ($G'$) obtainable from the peak frequency shift in an analogous manner to the longitudinal elastic modulus using Eqn. S4, S5, S9 and S10. The shear modulus is a stiffness coefficient ($c_{ij}$) that describes cases where the strain is pure shear. This means that the material's shape is changed but its density is not. It is the appropriate modulus for transverse sound waves. As with the longitudinal modulus, it can be different in different directions unless the material is isotropic. It will normally correspond to $c_{44}$, $c_{55}$, or $c_{66}$ of the stiffness tensor when measuring in a symmetry direction in a crystalline material. Typically two of these three coefficients show up as two distinct BLS peaks in a single measurement. These two peaks correspond to the two principal polarizations of the transverse phonon mode. The transverse modes are usually measured in distinct (90 degree) scattering geometries, although in anisotropic systems they can also be revealed in back-scattering measurements, provided that the wavevector is not along a direction of high symmetry. Because they are generally isotropic, amorphous materials will only have a single shear modulus.

**Linewidth**

The BLS linewidth ($\Gamma_B$), i.e. the Full Width at Half Maximum (FWHM) of the BLS peak, obtained from fitting a DHO or Lorentzian function (Eqns. S1 & S2), is fundamentally determined by the lifetime of the phonon--in an analogous manner to the peak width in a high energy physics experiment, with the (quasi-)particles in this case being the acoustic phonons. Given the phonons propagate at a presumed speed $V_q$, this lifetime is directly related to their spatial attenuation length. It is possible to also calculate a corresponding *loss modulus* that describes the dissipative properties of the sample (i.e. the corresponding imaginary part of the stiffness tensor components). In the case of longitudinal phonon modes this would be:

$$M'' = 2\pi \rho q^{-2} v_B \Gamma_B \qquad \textbf{Eqn. S12}$$

For a mechanically isotropic sample, $M''$ is the imaginary part of the *complex longitudinal modulus*: $M = M' + iM''$, and $i = \sqrt{-1}$. But as with the real part, $M''$ can in general take on different values in different directions (i.e. $M''_{ij}$). An analogous relation may be obtained from the BLS frequency shift and linewidth of the transverse acoustic modes, yielding the imaginary part of the shear modulus (G'').

For fluids it may sometimes also be desirable to express the dissipative properties in terms of a viscosity. For this the following relation may be used:

$$\Gamma_B = \frac{q^2}{2\pi\rho}\left[\frac{4}{3}\eta_S + \eta_B + \frac{\kappa}{C_P}(\gamma - 1)\right] \qquad \textbf{Eqn. S13}$$

Here $\eta_S$ is the *dynamic shear viscosity* common to rheology (but in this case that the probed MHz-GHz frequencies). $\eta_B$ is the *dynamic bulk viscosity* (also referred to as the second viscosity by e.g. Landau and Lifschitz), which is a distinct material property to the shear viscosity, that in practice can usually be neglected from fluid dynamics calculations, i.e.

Navier Stokes equation, on presumption of the incompressibility of flow. It however becomes relevant when there are e.g. bubbles, one considers microscopic instabilities, or of course for the propagation of longitudinal sound waves. $\kappa$ is the thermal conductivity of the sample, and $\gamma = C_P/C_V$ where $C_P$ ($C_V$) are its specific heat under constant pressure (volume). In water based matter one often can assume that $C_P \approx C_V$ such that $\gamma \approx 1$ and that the linewidth is directly proportional to the combination of the dynamic shear and bulk viscosity with no offset:

$$\Gamma_B \approx \frac{q^2 \eta_L}{2\pi\rho} \quad \textbf{Eqn. S14}$$

Where $\eta_L = (4/3)\eta_S + \eta_B$ is the *dynamic longitudinal viscosity* (sometimes also called the *effective* or *dilational* viscosity in the literature) that is relevant for describing the attenuation of longitudinal acoustic modes. $\eta_L$ can serve as a good reporting parameter insofar that in a Newtonian fluid under hydrodynamic conditions it will be independent of frequency and wavevector (compared to the loss modulus and acoustic attenuation, which would scale linearly and quadratically respectively with the probing wavevector--and thus also the refractive index). For some applications it may also be relevant to calculate the *kinematic longitudinal viscosity* $\mu_L$, which is defined as $\mu_L = \eta_L/\rho$.

For measurements of transverse phonons, the linewidth can analogously yield information on the imaginary part of the shear modulus ($G''$) and shear viscosity (Eqn. S12 and Eqn. S14 with the replacement $\eta_L \to \eta_S$ in the latter). The shear viscosity and moduli measured using BLS will be distinct from that measured using classical rheological techniques (shear rheology, etc. which measure at low shear rates) as it is typically at many orders of magnitude higher frequencies.

The above assumes that there are no extrinsic factors that cause broadening of the BLS peak (i.e. increases in the linewidth), and are defined for measurements at a single scattering wavevector. In practice numerous factors will also cause broadening of the BLS scattering peak and over estimations of the linewidth. These include measurements from a distribution of scattering angles (finite NA), material heterogeneities in the probing volume (that are larger than the characteristic length scales of the probed phonons), multiple scattering (most relevant for non-transparent samples or deep tissue imaging), and ultimately the usually non-negligible spectral Instrument Response Function (IRF) of the spectrometer. In addition when the probed phonon frequency is in the vicinity of a structural relaxation process with characteristic time(s) $\tau$, namely $\nu_B \sim \tau^{-1}$, as may happen to be the case in soft-matter/gels, the functional dependence of the BLS scattering peak will undergo significant broadening and deviate from a simple DHO.

While a spectral deconvolution may suffice in correcting for the IRF and finite-NA broadening, accounting for the other factors analytically or even numerically is less trivial, requiring a priori information on the sample that is usually not available. To this end a experiment/sample-specific semi-quantitative (or atleast qualitative) assessment of their expected relative contributions should be made, and caution should be exercised when presenting absolute values of linewidth-derived parameters.

**Loss tangent**

This can serve as a useful metric of changes in the viscoelastic properties owing to it being independent of changes in the mass density (which are challenging to measure on the microscopic scale in complex biological samples). It is defined, for the longitudinal modes, as:

$$\tan \delta = \Gamma_B/\nu_B = M''/M'  \qquad \text{Eqn. S15}$$

An analogous quantity may also be defined for the transverse (shear) modes. It is sometimes stated that $\tan \delta$ is independent of the refractive index $n$. This is only true if both $M''$ and $M'$ are assumed to be independent of the phonon wavevector probed $q$ (Eqn. S5) (or have the exact same functional dependence thereon). While potentially valid in hard dehydrated solids, this is not the case in the hydrodynamic picture, where $\Gamma_B \propto q^2$ (Eqn. S13) and $\nu_B \propto q$ (Eqn. S4), such that $\tan \delta \propto q \propto n$. In the vicinity of structural and phase transitions this assumption can be assumed to break down even more dramatically.

**Relation to other moduli**

In an isotropic material, the longitudinal elastic modulus is related to the shear and bulk elastic modulus via:

$$M' = (4/3)G' + K'  \qquad \text{Eqn. S16}$$

This is true also for the imaginary component (loss moduli), with the longitudinal viscosity being defined analogously: $\eta_L = (4/3)\eta_S + \eta_B$ (see Eqn. S13 and S14).

In a linear, isotropic material the longitudinal modulus (M), Young's modulus ($E$), Shear Modulus ($G$) and bulk modulus ($K$) are related via the Poisson ratio ($\sigma$):

$$M = \frac{E(1-\sigma)}{(1+\sigma)(1-2\sigma)}; \quad E = \frac{9KG}{(3K+G)}; \quad G = \frac{E}{2(1+\sigma)}; \quad K = \frac{E}{3(1-2\sigma)}$$

$$M = \frac{1-\sigma}{(1+\sigma)(1-2\sigma)}E = \frac{2(1-\sigma)}{(1-2\sigma)}G = \frac{3(1-\sigma)}{1+\sigma}K  \qquad \text{Eqn. S16}$$

These equations are only valid $-1 < \sigma < 0.5$, with the material becoming inherently unstable outside of this range. In practice, their relevance will also already break down when E/G and M/K become significantly different as will be the case when $\sigma$ approaches 0.5. Here one would need to know $\sigma$ to an increasingly higher accuracy (to an infinite accuracy in the limiting case of $\sigma = 0.5$). It should be kept in mind that the Poisson ratio is also a frequency dependent quantity, and can not generally be assumed to have the same values at MHz-GHz as it does at low frequencies.

While Eqn.S15 and S16 may be used to convert between moduli (assuming one has knowledge of the Poisson's ratio at the relevant MHz-GHz frequencies), the obtained e.g. $E$ or $G$ can in the general case not simply be compared to measurements obtained from techniques probing at quasi-static frequencies (AFM, shear rheology, etc.) due to the many orders of magnitude differences in frequency, and often unknown mechanical relaxation spectrum between these frequencies. For specific sample types it may however be possible to realize significant and useful empirical correlations between the different moduli, and even the same or different moduli at different frequencies.

An additional caveat for calculating e.g. $E$ from the BLS measured $M$ and $G$ that may become relevant, is that the BLS measured $M$ and $G$ may in themselves be at quite different frequencies--namely the BLS frequency shifts measured for the longitudinal and transverse phonons may be a couple of orders of magnitude apart. Eqn.s S15 and S16 on the other hand assume that all quantities are at the same frequency.

Much of the above is based upon the assumption of hydrodynamic behaviour and that there are no significant phase or structural transitions at or near the probed BLS frequencies. This may or may not be the case in biorelevant matter, and this should always be kept in mind.

Finally, a difference in the BLS measurements and perturbation-based measurements of these these moduli may also be due to the latter usually measuring elastic moduli by applying a finite-strain (obtaining the so-called *engineering* stress-strain), whereas BLS probing the isentropic *true* stress-strain.

**Calculation of stiffness tensor**

It is in principle possible to determine the complete stiffness tensor by performing BLS measurements of the sample with different scattering wavevectors. Measurements of an anisotropic sample in an arbitrary direction that does not correspond to a symmetry axis, will yield a possibly complex algebraic combination of stiffness tensor coefficients (which can be determined by Eqn. S8). Such measurements are relatively uncommon except in cases where the experimenter is explicitly trying to extract the values of all stiffness tensor elements including off-diagonal elements such as $c_{12}$. Most measurements are usually performed in high symmetry directions, where the effective stiffness corresponds to just a single coefficient from the stiffness tensor, such as $c_{11}$, $c_{22}$, or $c_{33}$ for a longitudinal mode or $c_{44}$, $c_{55}$, or $c_{66}$ for a transverse mode (see Figure S1).

$$\begin{pmatrix}\sigma_{xx}\\ \sigma_{yy}\\ \sigma_{zz}\\ \hdashline \sigma_{yz}\\ \sigma_{zx}\\ \sigma_{xy}\end{pmatrix} = \begin{pmatrix}c_{11} & c_{12} & c_{13} & \vdots & c_{14} & c_{15} & c_{16}\\ c_{21} & c_{22} & c_{23} & \vdots & c_{24} & c_{25} & c_{26}\\ c_{31} & c_{32} & c_{33} & \vdots & c_{34} & c_{35} & c_{36}\\ \hdashline c_{41} & c_{42} & c_{43} & \vdots & c_{44} & c_{45} & c_{46}\\ c_{51} & c_{52} & c_{53} & \vdots & c_{54} & c_{55} & c_{56}\\ c_{61} & c_{62} & c_{63} & \vdots & c_{64} & c_{65} & c_{66}\end{pmatrix}\begin{pmatrix}\varepsilon_{xx}\\ \varepsilon_{yy}\\ \varepsilon_{zz}\\ \hdashline \varepsilon_{yz}\\ \varepsilon_{zx}\\ \varepsilon_{xy}\end{pmatrix}$$

(longitudinal columns: $c_{11}, c_{22}, c_{33}$; shear diagonal: $c_{44}, c_{55}, c_{66}$; strain components: longitudinal $\varepsilon_{xx},\varepsilon_{yy},\varepsilon_{zz}$ and shear $\varepsilon_{yz},\varepsilon_{zx},\varepsilon_{xy}$.)

**Figure S1.** BLS in a symmetric direction measures the blue (longitudinal modes) or red (transverse modes). All other stiffnesses are of mixed quasi-transverse / quasi-longitudinal nature and are found by fitting the BLS signal to the Christoffel Equations.

To measure all the stiffness tensor elements requires measuring the sound speeds along different directions which will correspond to the different stiffnesses ($c_{11}$, $c_{22}$, etc.) using a combination of different scattering geometries and rotations of the sample. Some common scattering geometries are shown in Supplementary Table 1. The BLS experiment should be altered depending on the symmetry of the sample. A truly isotropic material only has two independent stiffness coefficients ($c_{11}$ the longitudinal component, and $c_{44}$ the transverse component) and these are easily measured, without needing to know the refractive index, by using an equal angle scattering geometry (90a, see Supplementary Table 1). This can however be complicated in cells and other small materials. A single spectrum in an equal angle geometry will give a longitudinal mode, which gives $c_{11}$, and a transverse mode, which gives $c_{44}$. This gives all the stiffness tensor elements for an isotropic material.

| Geometry | Schematic | Equation |
|---|---|---|
| **180** | 180a 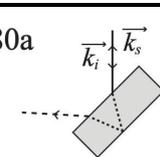 | $V_{180} = \dfrac{\Delta v_{180}\, \lambda_o}{2n}$ |
| **90n** | 90n 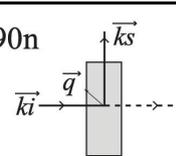 | $V_{90n} = \dfrac{\Delta v_{90n}\, \lambda_o}{n\sqrt{2}}$ |
| **90a Equal-Angle** | 90a 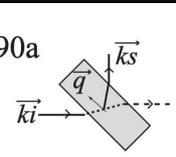 | $V_{90a} = \dfrac{\Delta v_{90a}\, \lambda_o}{\sqrt{2}}$ |
| **90r** | 90r 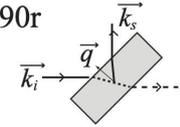 | $V_{90r} = \dfrac{\Delta v_{90r}\, \lambda_o}{\sqrt{4n^2 - 2}}$ |
| **Platelet (60°)** | Platelet 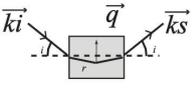 | $V_{60} = \Delta v_{60}\, \lambda_o$ |

**Supplementary Table 1:** Common BLS scattering geometries used for determining the diagonal stiffness tensor coefficients in materials of known symmetries. $k_i(k_s)$ are incident and scattering wavevector, $q$ is the scattering wavevector, and $\Delta v$ is the measured frequency shift.

For some applications, one wants to know other elastic properties such as the Young's modulus ($E$), the bulk modulus ($K$), the Poisson's ratio ($\sigma$), or the shear modulus ($G$), averaged over crystal orientations. These can all be derived from the stiffness coefficients $c_{ij}$. For an isotropic material we have three distinct stiffness coefficients $c_{11}$, $c_{12}$, and $c_{44}$, but only two are independent because of the constraint $c_{44} = (c_{11} - c_{12})/2$. The different moduli can then be found from:

$$E = \frac{3c_{12}c_{44} + 2c_{44}^2}{c_{12} + c_{44}} = \frac{(c_{11} - c_{12})(c_{11} + 2c_{12})}{c_{11} + c_{12}}$$

$$G = c_{44}$$

$$K = \frac{1}{3}(c_{11} + 2c_{12})$$

$$\sigma = \frac{c_{12}}{c_{11} + c_{12}} \qquad \text{Eqn. S17}$$

As also apparent from Eqn. S17, these equations show that the longitudinal modulus $c_{11}$ is not the same as the Young's modulus but is some combination of the Young's modulus and the Poisson's ratio – expressed in this case as combinations of stiffness tensor elements.

Determining the entire stiffness tensor in a more complex anisotropic sample requires measurements in several directions of the material including off-axis or off-symmetry directions. The experimental design must ensure that the measurements are sensitive to all the stiffness tensor elements and not just the ones along the main diagonal. This means measuring not just pure longitudinal and transverse modes but also mixed modes, so-called *quasi-longitudinal* and *quasi-transverse* mode. BLS frequency shifts are measured as the material is rotated from one on-axis direction to the next, thus rotating the direction of the phonon wave-vector in the material. With enough such measurements, one can numerically fit the sound velocities to the Christoffel equations and obtain all the independent stiffness tensor elements (and, in some cases, the refractive index). Measured values are fit, usually via a linear least square model, to solve for the individual coefficients. Some common stiffness tensor symmetries that have been used to describe biological samples are shown in Figure S2.

An analytic form of the Christoffel equations for a *cubic symmetry* can be obtained. When the wavevector, $k$, lies in the (001) plane at an angle $\phi$ relative to the x-axis these are given by:

$$V_L = \sqrt{\frac{\frac{1}{2}\{(c_{11} + c_{44}) \pm [(c_{11} - c_{44})^2 - 4A(c_{11} + c_{12})\cos^2\phi \sin^2\phi]^{1/2}\}}{\rho}}$$

$$V_{T_2} = \sqrt{\frac{\frac{1}{2}\{(c_{11} + c_{44}) \pm [(c_{11} - c_{44})^2 - 4A(c_{11} + c_{12})\cos^2\phi \sin^2\phi]^{\frac{1}{2}}\}}{\rho}}$$

$$V_{T_1} = \sqrt{\frac{c_{44}}{\rho}} \qquad \text{Eqn. S19}$$

Isotropic

$$\begin{pmatrix} c_{11} & c_{12} & c_{12} & 0 & 0 & 0 \\ c_{12} & c_{11} & c_{12} & 0 & 0 & 0 \\ c_{12} & c_{12} & c_{11} & 0 & 0 & 0 \\ 0 & 0 & 0 & \frac{1}{2}(c_{11}-c_{12}) & 0 & 0 \\ 0 & 0 & 0 & 0 & \frac{1}{2}(c_{11}-c_{12}) & 0 \\ 0 & 0 & 0 & 0 & 0 & \frac{1}{2}(c_{11}-c_{12}) \end{pmatrix}$$

Cubic

$$\begin{pmatrix} c_{11} & c_{12} & c_{12} & 0 & 0 & 0 \\ c_{12} & c_{11} & c_{12} & 0 & 0 & 0 \\ c_{12} & c_{12} & c_{11} & 0 & 0 & 0 \\ 0 & 0 & 0 & c_{44} & 0 & 0 \\ 0 & 0 & 0 & 0 & c_{44} & 0 \\ 0 & 0 & 0 & 0 & 0 & c_{44} \end{pmatrix}$$

Hexagonal

$$\begin{pmatrix} c_{11} & c_{12} & c_{13} & 0 & 0 & 0 \\ c_{12} & c_{11} & c_{13} & 0 & 0 & 0 \\ c_{13} & c_{13} & c_{11} & 0 & 0 & 0 \\ 0 & 0 & 0 & c_{44} & 0 & 0 \\ 0 & 0 & 0 & 0 & c_{44} & 0 \\ 0 & 0 & 0 & 0 & 0 & \frac{1}{2}(c_{11}-c_{12}) \end{pmatrix}$$

**Figure S2.** Three of the most common stiffness tensor symmetries used to describe biological systems measured with BLS. Note: In the isotropic, cubic and hexagonal cases there are only 2, 3 and 4 independent components respectively.

When on the other hand the wavevector, $k$, is in the (011) plane at an angle $\phi$ relative to the x-axis these will be given by:

$$V_L = \sqrt{\frac{\frac{1}{4}\{(c_{11}+c_{12}+4c_{44}) + A_1 cos^2\phi \pm [(c_{11}+c_{12})^2 - A_1(6c_{11}+14c_{12}+8c_{44})cos^2\phi + A_1(9c_{11}+15c_{12}+6c_{44})cos^4\phi]^{1/2}}{\rho}}$$

$$V_{T_1} = \sqrt{\frac{\frac{1}{4}\{(c_{11}+c_{12}+4c_{44}) + A_1 cos^2\phi \pm [(c_{11}+c_{12})^2 - A_1(6c_{11}+14c_{12}+8c_{44})cos^2\phi + A_1(9c_{11}+15c_{12}+6c_{44})cos^4\phi]^{1/2}}{\rho}}$$

$$V_{T_2} = \sqrt{\frac{\frac{1}{2}[(c_{11}-c_{12})-A_2 cos^2\phi]}{\rho}} \quad \textbf{Eqn. S20}$$

where $A_1 = c_{11} - c_{12} - 2c_{44}$ and $A_2 = c_{11} - c_{12} - 2c_{44}$. In each case one fits angle ($\phi$) resolved measurements to extract the different unknown stiffness tensor coefficients. The bulk, Young's and shear elastic moduli can be calculated from:

$$K = \frac{(c_{11}+2c_{12})}{3} \quad \textbf{Eqn. S21a}$$

$$E = \frac{(c_{11}-c_{12})(c_{11}+2c_{12})}{(c_{11}+c_{12})} \quad \textbf{Eqn. S21b}$$

$$G = c_{44} = \frac{(c_{11}-c_{12})}{2} \quad \textbf{Eqn. S21c}$$

The analytical form of the Christoffel equations for a system with *hexagonal symmetry* are given by:

$$V_L = \frac{c_{11}sin^2\phi + c_{33}cos^2\phi + c_{44} \pm \sqrt{[(c_{11}-c_{44})sin^2\phi + (c_{44}-c_{33})cos^2\phi]^2 + 4(c_{13}+c_{44})^2 sin^2\phi cos^2\phi}}{2\rho}$$

$$V_{T_1} = \frac{c_{11}sin^2\phi + c_{33}cos^2\phi + c_{44} \pm \sqrt{[(c_{11}-c_{44})sin^2\phi + (c_{44}-c_{33})cos^2\phi]^2 + 4(c_{13}+c_{44})^2 sin^2\phi cos^2\phi}}{2\rho}$$

$$V_{T_2} = \sqrt{\frac{c_{66}sin^2\phi + c_{44}cos^2\phi}{\rho}} \quad \textbf{Eqn. S22}$$

Here the bulk, Young's and shear elastic moduli will be given by:

$$K = \frac{-2c_{13}^2 + (c_{11}+c_{12})c_{33}}{c_{11}+c_{12}-4c_{13}+2c_{33}} \quad \textbf{Eqn. S23a}$$

$$E_\parallel = c_{33} - \frac{2c_{13}^2}{c_{11}+c_{12}} \qquad E_\perp = \frac{(c_{11}-c_{12})[c_{33}(c_{11}+c_{12})-2c_{13}^2]}{c_{11}c_{33}-c_{13}^2} \quad \textbf{Eqn. S23b}$$

$$G_1 = c_{44} \qquad G_2 = c_{66} \qquad \textbf{Eqn. S23c}$$

$$\sigma_{12} = \frac{c_{33}c_{12} - c_{13}^2}{c_{33}c_{11} - c_{13}^2} \qquad \sigma_{13} = \frac{c_{13}}{c_{11} + c_{12}} \qquad \textbf{Eqn. S23d}$$

**Registration to accepted/standard values**

Different BLS instruments operate at different wavelengths and there are various extrinsic factors that can affect the BLS frequency shift ($\nu_B$) and linewidth ($\Gamma_B$). It has previously been proposed that rather than simply reporting $\nu_B$ and $\Gamma_B$ it may be desirable to report $\underline{\nu_B} = \nu_B/\nu_B^w - 1$ and $\underline{\Gamma_B} = \Gamma_B/\Gamma_B^w - 1$, where $\nu_B^w$ and $\Gamma_B^w$ are the BLS frequency shift and linewidth of water measured using the same instrument, at the same temperatures and experimental conditions. The intention thereof was to allow for easy comparison of measurements from different instruments with different probing wavelengths, and to some degree also account for different artifacts that would perturb measurements of the probed sample in a similar way as water. These however do not account for constant systematic offsets or spectral broadening resulting from spectrometer design and measurement geometry. We thereby propose the following correction.

For $\nu_B$ a systematic offset can be accounted for by comparing the measured frequency shift for *pure* water (ideally from an injection grade ampoule, else double or triple distilled) $\nu_B^{(w)}$ measured with the same scattering geometry (with effective scattering wavevector $q^{(w)}$, Eqn. S5) as the sample, with an accepted value for water $\nu_B^{(w,0)}$ at the same temperature ($T$). We propose the following correction for the BLS frequency shift $\nu_B^{(a)}$ measured for sample "$a$":

$$\nu'^{(a)}_B(T) = \nu_B^{(a)}(T) - \nu_B^{(w)}(T) + \left(q^{(w)}/q^{(w,0)}\right)\nu_B^{(w,0)}(T) \qquad \textbf{Eqn. S24}$$

Where $\nu_B^{(w,0)}$ is and accepted standard value of water, and $q^{(w,0)}$ the scattering wavevector of the standard value measurement. From this it follows that the corrected hypersonic speed $V'^{(a)}(T)$ will be given by:

$$V'^{(a)}(T) = (2\pi/q^{(a)})\left[\nu_B^{(a)}(T) - \nu_B^{(w)}(T) + \left(q^{(w)}/q^{(w,0)}\right)\nu_B^{(w,0)}(T)\right] \qquad \textbf{Eqn. S25}$$

and the change in the hypersonic speed relative to the corrected value for water can compactly be written as:

$$\frac{V'^{(a)}(T)}{V'^{(w)}(T)} = \frac{q^{(w,0)}}{q^{(a)}}\left[\frac{\nu_B^{(a)}(T) - \nu_B^{(w)}(T)}{\nu_B^{(w,0)}(T)}\right] + \frac{q^{(w)}}{q^{(a)}} \qquad \textbf{Eqn. S26}$$

An analogous correction can be applied to the linewidth:

$$\Gamma'^{(a)}_B(T) = \Gamma^{(a)}_B(T) - \Gamma^{(w)}_B(T) + \left(q^{(w)}/q^{(w,0)}\right)^2 \Gamma^{(w,0)}_B(T) \qquad \text{Eqn. S27}$$

Where $\Gamma^{(w,0)}_B$ is the accepted standard value of water. It follows that the corrected kinematic viscosity ($\mu_L$) will be given by:

$$\mu'^{(a)}_L(T) = (2\pi/q^{(a)\,2})\left[\Gamma^{(a)}_B(T) - \Gamma^{(w)}_B(T) + \left(q^{(w)}/q^{(w,0)}\right)^2 \Gamma^{(w,0)}_B(T)\right] \qquad \text{Eqn. S28}$$

and the relative difference of the corrected kinematic viscosity to that of water will be:

$$\frac{\mu'^{(a)}_L(T)}{\mu'^{(w)}_L(T)} = \left(\frac{q^{(w,0)}}{q^{(a)}}\right)^2 \left[\frac{\Gamma^{(a)}_B(T) - \Gamma^{(w)}_B(T)}{\Gamma^{(w,0)}_B(T)}\right] + \left(\frac{q^{(w)}}{q^{(a)}}\right)^2 \qquad \text{Eqn. S29}$$

An interrelated table of *standard values* for $v^{(w,0)}_B(T)$, $\Gamma^{(w,0)}_B(T)$ and $q^{(w,0)}$ that may be used, are provided at DOI:10.6084/m9.figshare.27794913. These were obtained from deconvolved TFP measurements on injection grade water ampoules in laboratories in Vienna and Hannover. They were performed in ≤1°C temperature steps over the range 20-40°C, and throughout showed agreement of better than 0.1% with each other for both the frequency shift and linewidth. The acoustic speed derived from these values agreed with the accepted acoustic velocity in pure water to within 0.2% over the temperature range 25-35°C for which the latter was available.

**Relation between refractive index and mass density**

The *Clausius–Mossotti equation* or equivalently the *Lorenz-Lorentz equation*, relates the (macroscopic) optical properties of a material to its microscopic physical properties (molecular polarizability and molecular density), and has been shown to be applicable in diverse solid materials. In particular it predicts a relation between the refractive index ($n$) and the mass density ($\rho$):

$$\frac{(n^2-1)}{(n^2+2)} = \left(\frac{4\pi}{3}\right)N\alpha = \left(\frac{4\pi}{3}\right)\left(\frac{\rho}{M}\right)\alpha \qquad \text{Eqn. S30}$$

where $N$ is the number density of molecules, $M$ is their molecular weight, and $\alpha$ is the electronic polarizability. When the electronic polarizability or the number density of molecules is small (as $N\alpha \to 0$), this reduces to: $n^2 - 1 \approx 4\pi M^{-1}\alpha\rho$ and further to $n - 1 \approx 2\pi M^{-1}\alpha\rho$. These approximations have been found to be valid for gases under normal pressure. As can be seen none of these relations directly imply that $n^2/\rho$ is constant (yet alone material independent) as would be desired for the BLS frequency shift to be related directly to the elastic modulus (Eqn. S10). As such independent measurements of both the refractive index $n$ and mass density $\rho$ are in general required to obtain accurate prediction of viscoelastic moduli from BLS data.

## Supplementary References